\documentclass[twocolumn,floatfix,showpacs,amsmath,amssymb,pra]{revtex4}
\usepackage{times}
\usepackage{graphicx}
\usepackage{dcolumn}
\usepackage{bm}

\makeatletter
\def\footnotelabel{%
  \edef\@currentlabel{\p@footnote\thefootnote}%
  \label}
\makeatother

\def\vecgr#1{\mathchoice{\mbox{\boldmath$\mathrm\displaystyle#1$}}
{\mbox{\boldmath$\mathrm\textstyle#1$}}
{\mbox{\boldmath$\mathrm\scriptstyle#1$}}
{\mbox{\boldmath$\mathrm\scriptscriptstyle#1$}}}

\def\vec#1{\mathchoice{\mathrm{\mathbf{\displaystyle#1}}}
{\mathrm{\mathbf{\textstyle#1}}}
{\mathrm{\mathbf{\scriptstyle#1}}}
{\mathrm{\mathbf{\scriptscriptstyle#1}}}}
\newcommand{\bi}{\begin{itemize}}
\newcommand{\ei}{\end{itemize}}

\newcommand{\be}{\begin{equation}}
\newcommand{\ee}{\end{equation}}
\newcommand{\ba}{\begin{array}}
\newcommand{\ea}{\end{array}}
\newcommand{\bea}{\begin{eqnarray}}
\newcommand{\eea}{\end{eqnarray}}

\newcommand{\bc}{\begin{center}}
\newcommand{\ec}{\end{center}}
\newcommand{\bslide}{\begin{slide}}
\newcommand{\eslide}{\end{slide}}

\newsavebox{\TRS}
\sbox{\TRS}{\hspace{.5em} = \hspace{-1.8em}
                 \raisebox{1ex}{\mbox{\scriptsize TRS}} }

\newsavebox{\defgleich}
\sbox{\defgleich}{\ :=\ }

\newsavebox{\LSIM}
\sbox{\LSIM}{\raisebox{-1ex}{$\ \stackrel{\textstyle<}{\sim}\ $}}

\newsavebox{\GSIM}
\sbox{\GSIM}{\raisebox{-1ex}{$\ \stackrel{\textstyle>}{\sim}\ $}}

\newcounter{saveeqn}

\newcommand{\ptd}{\partial}

\newcommand{\Tr}{\mbox{Tr$\,$}}

\newcommand{\intdd}[1]{\int\,d^3\!#1\,}

\newcommand{\intddz}[2]{\int\,d^3\!#1\,d^3\!#2\,}

\newcommand{\cC}{{\cal C}}
\newcommand{\cD}{{\cal D}}

\newcommand{\cN}{{\cal N}}
\newcommand{\cO}{{\cal O}}

\newcommand{\cT}{{\cal T}}

\newcommand{\nabv}{\vecgr{\nabla}}

\newcommand{\pv}{\vec{p}}

\newcommand{\xv}{\vec{x}}

\newcommand{\yv}{\vec{y}}

\newcommand{\zv}{\vec{z}}

\newcommand{\tr}{\hbox{tr}}

\begin{document}

\title{Nonperturbative dynamical many-body theory of a Bose-Einstein condensate}
\author{Thomas Gasenzer}
\thanks{email:T.Gasenzer@thphys.uni-heidelberg.de}
\author{J\"urgen Berges}
\thanks{email:J.Berges@thphys.uni-heidelberg.de}
\author{Michael G. Schmidt}
\thanks{email:M.G.Schmidt@thphys.uni-heidelberg.de}
\author{Marcos Seco}
\thanks{email:M.Seco@thphys.uni-heidelberg.de}
\affiliation{Institut f\"ur Theoretische Physik, Universit\"at Heidelberg, Philosophenweg 16, 
69120 Heidelberg, Germany}
\date{\today}

\begin{abstract} 
\noindent
A dynamical many-body theory is presented which systematically extends beyond mean-field and perturbative quantum-field theoretical procedures.
It allows us to study the dynamics of strongly interacting quantum-degenerate atomic gases.
The non-perturbative approximation scheme is based on a systematic expansion of the two-particle irreducible effective action in powers of the inverse number of field components. 
This yields dynamic equations which contain direct scattering, memory and ``off-shell'' effects that are not captured by the Gross-Pitaevskii equation. 
This is relevant to account for the dynamics of, e.g., strongly interacting quantum gases atoms near a scattering resonance, or of one-dimensional Bose gases in the Tonks-Girardeau regime. 
We apply the theory to a homogeneous ultracold Bose gas in one spatial dimension. 
Considering the time evolution of an initial state far from equilibrium we show that it quickly evolves to a non-equilibrium quasistationary state and discuss the possibility to attribute an effective temperature to it. 
The approach to thermal equilibrium is found to be extremely slow.
\end{abstract}
\pacs{03.75.Kk, 03.75.Nt, 05.30.-d, 05.70.Ln, 11.15.Pg\hfill HD--THEP--05--13}
\maketitle
\section{Introduction}
\label{sec:intro}
Since the pioneering achievement of Bose-Einstein condensation in dilute alkali-metal gases much effort has been made to extend this success to a wider class of atomic and molecular species as well as to a wealth of different trapping geometries.
In particular, the production of quantum degenerate ensembles of molecules \cite{GossLevi2000a,Williams2000a} promises a wide spectrum of important applications ranging from ultra-precise molecular spectroscopy and cold collision studies \cite{Weiner1999a} to ``superchemical'' reactions \cite{Heinzen2000a} and the investigation of the crossover from a Bose-Einstein condensate (BEC) of molecules to Bardeen-Cooper-Schrieffer (BCS) correlated pairs in Fermi gases \cite{Randeria1995a}.
Furthermore, (quasi) one- and two-dimensional traps \cite{Schmiedmayer2000a,Pitaevskii2003a} as well as optical lattices \cite{Jaksch1998a,Bloch2004a} allow us to realize strongly correlated many-body states of atoms reminiscent of similar phenomena in condensed matter systems.

Zero-energy scattering resonances, particularly the so called magnetic Feshbach resonances \cite{Stwalley1976b,Tiesinga1992a,Tiesinga1993a,Burnett1998a,Mies2000a} so far have played a leading r\^{o}le in the creation of strong interactions in degenerate atomic quantum gases.
Near a Feshbach resonance, the scattering of, e.g., a pair of Bose-condensed atoms, whose relative energy is very close to zero, can be described by a strongly enhanced $s$-wave scattering length $a$.
Present-day experimental techniques allow for resonance-enhanced scattering lengths larger than the mean interatomic distance $n^{-1/3}$ in the gas.
As a consequence, the diluteness parameter $\eta=na^3$ is larger than one.
The Bose-Einstein condensate is no longer in the collisionless regime, it represents a strongly interacting system.

Feshbach resonances also play an important r\^{o}le in the physics of ultracold, degenerate Fermi gases, where they allow us to induce a transition from a phase of Bose-Einstein-condensed diatomic molecules to a superfluid phase of BCS-type where the Fermion pairs are correlated over large distances \cite{Regal2004b,Zwierlein2004a,Bartenstein2004a}.
In the transition region, the interaction is strong and the correlation length easily exceeds the mean interatomic distance.

A strongly interacting system can also be obtained without the requirement of a scattering resonance:
In a one-dimensional trap, the gas enters the so-called Tonks-Girardeau regime, if the dimensionless interaction parameter $\gamma=g_\mathrm{1D}m/(\hbar^2n)$ is much larger than one \cite{Tonks1936a,Girardeau1960a,Paredes2004a}. 
Here, $g_\mathrm{1D}$ is the coupling parameter of the one-dimensional gas, e.g., $g_\mathrm{1D}=2\hbar^2a/(ml_\perp^2)$ for a cylindrical trap with transverse harmonic oscillator length $l_\perp$.
In the Tonks-Girardeau limit $\gamma\to\infty$ the atoms can no longer pass
each other and behave in many respects like a one-dimensional ideal Fermi
gas \cite{Pitaevskii2003a}. 

In an optical lattice, strong effective interactions can be induced by suppressing the hopping between adjacent lattice sites and thus increasing the weight of the interaction relative to the kinetic energy \cite{Jaksch1998a,vanOosten2001a}.
This leads, in the limit of near-zero hopping or strong interactions, to a Mott insulating state \cite{Greiner2002a}.

Conventionally, the dynamics of many-body systems of suffiently weakly interacting particles is described using perturbative approximation schemes for the exact quantum field theoretical many-body equations of motion.
Such schemes are in principle based on expansions in terms of powers of some dimensionless parameter, like $\sqrt{\eta}$, which measures the binary interaction strength.
It is clear that for strongly interacting systems for which the relevant
parameter is no longer small, a perturbative approach must eventually fail.

A particular challenge represent systems far from equilibrium. 
For instance, the sudden change of an external magnetic field for atomic quantum gases near a Feshbach resonance can lead to dramatic non-equilibrium phenomena \cite{Roberts2001a,Donley2002a}. 
If the magnetic field is modified on time scales smaller than a typical collisional duration, the system is driven far from thermal equilibrium. 
Standard approaches based on small deviations from equilibrium, such as linear response theory, or kinetic descriptions requiring a sufficient homogeneity in time, are not applicable.
Another example is the formation of a Bose-Einstein condensate through controlled evaporative cooling of an ultracold atomic gas, an every-day experimental procedure, which has been studied in great detail (cf.~e.g.~Refs.~\cite{Kagan1992a,Stoof1991a,Stoof1997a,Gardiner1998b,Davis2002a,Davis2002b}), a complete theoretical description of which is, however, still lacking \cite{Kohl2002a}.  

In this article a dynamical many body theory is presented which systematically extends beyond mean-field and perturbative quantum-field theoretical approximation schemes.
The non-perturbative approach is based on a systematic expansion of the two-particle irreducible (2PI) effective action in powers of the inverse
number of field components ${\cal N}$ \cite{Cornwall1974a,Berges2002a,Aarts2002b}. 
The 2PI $1/{\cal N}$ expansion to next-to-leading order yields dynamic equations which contain direct scattering, memory and ``off-shell'' effects. 
It allows to describe far-from-equilibrium dynamics as well as the late-time approach to quantum thermal equilibrium.

Recently, these methods have allowed important progress in describing the dynamics of strongly interacting relativistic systems far from thermal equilibrium for bosonic \cite{Berges2002a,Berges2003b,Mihaila2001a,Cooper2003a,Arrizabalaga2004a} as well as fermionic degrees of freedom \cite{Berges2003a,Berges2004b}. 
Our aim is to employ the 2PI effective action for ultracold quantum gases and to numerically solve the 2PI $1/{\cal N}$ expansion to next-to-leading order. 
This is exemplified for a homogeneous ultracold Bose gas in one spatial dimension. 
We compute the time evolution of an initial state which is far from thermal equilibrium. 
After a characteristic short-time scale it is found to be driven to a quasistationary state.
However, the system is still far from equilibrium and the thermal equilibration  time can exceed the early-time scale by orders of magnitude. 
In particular, a unique temperature can not be attributed to the quasistationary state.
This is important in view of experiments with one-dimensional traps, where the longest accessible times may be too short to see complete thermalization.

We emphasize that mean-field approximations fail to describe the dynamics even qualitatively. 
Similarly, standard kinetic descriptions based on two-to-two collisions give a trivial (constant) dynamics because of phase space restrictions in one spatial dimension. 
To our knowledge this is the first time that the one-dimensional dynamics has been described from a full systematic 2PI $1/{\cal N}$ expansion in a non-relativistic quantum field theory. 
The 2PI effective action approach in this context has been discussed previously using an additional weak-coupling expansion and solved for a Bose gas in a lattice in one spatial dimension \cite{Rey2004a}.
See also Ref. \cite{Andersen2004a} for a discussion of the 2PI (or $\Phi$-derivable) approach to the theory of weakly interacting Bose gases.   
The connection to kinetic theory (cf., e.g., Refs.~\cite{Proukakis1996a,Shi1998a,Giorgini1998a,Walser1999a,Bhongale2002a,Davis2000a,Stoof1999a,Imamovic-Tomasovic2001a,Boyanovsky2002a} and references therein) has been discussed in
Refs.~\cite{Rey2005a,Baier2005a}, cf.~also
Refs.~\cite{Calzetta1988a,Ivanov1999a,Prokopec2004a,Prokopec2004b,Konstandin2005a,Konstandin2005b} in the context of elementary particle physics and cosmology. 
The 2PI $1/{\cal N}$ expansion has also been successfully applied to compute critical exponents in thermal equilibrium near the second-order phase transition of a model in the same universality class \cite{Alford2004a}.
The good convergence properties of the expansion for small values of ${\cal N} > 1$ have also been observed in the context of non-equilibrium classical statistical field theories \cite{Aarts2002a}.

Ultracold atomic Bose gases in one- and two-dimensional trapping configurations, due to their specific quantum statistical properties, have been studied in great detail, using both perturbative and non-perturbative approaches. 
In lower dimensional systems, the regime of applicability of perturbation theory (cf., e.g., \cite{Petrov2000a,Petrov2000b}) is reduced as compared to three-dimensional gases since fluctuations, in particular of the phase, play a stronger r\^{o}le such that non-perturbative descriptions are required.
Stationary systems have been studied using, e.g., renormalization group \cite{Andersen2002b,AlKhawaja2002a} or Monte-Carlo techniques \cite{Astrakharchik2004a,Pilati2005a}.

Our article is organized as follows:
In the remainder of this chapter we introduce the basic principles and ideas underlying the non-perturbative approximation scheme.
In Chapter \ref{sec:EffActionApproach} the 2PI effective action is defined, and the (exact) dynamic equations are deduced from it.
We then show, in Chapter \ref{sec:PT}, that the well-known dynamical Hartree-Fock-Bogoliubov equations of motion result from a single diagram in a loop expansion of the 2PI effective action.
In Chapter \ref{sec:Nonpert}, we introduce in detail the non-perturbative approximation scheme on the basis of a $1/{\cal N}$ expansion of the 2PI effective action, where $\cal N$ is the number of field components in a subspace where the action is invariant under $O({\cal N})$ rotations.
For the simplest case of a single complex Bose-field, we have ${\cal N}=2$.
Before we draw our conclusions in Chapter \ref{sec:Concl} we apply, in Chapter \ref{sec:1DDynamics}, the many-body dynamic equations, to next-to-leading order in the $1/{\cal N}$ expansion, to a uniform ultracold Bose gas in one spatial dimension.

\subsection{Non-perturbative approximations out of equilibrium}
\label{sec:PTvsNPT}

The systems of identical bosons to be considered in this article are described
by the Hamiltonian  
\begin{align}
\label{eq:MBHamiltonian}
 &H
 =\intdd{x}\hat\Psi^\dagger(\xv)H_\mathrm{1B}(\xv)\hat\Psi(\xv)
 \nonumber\\
 &\qquad
 +\ \frac{1}{2}\intddz{x}{y}\,
     \hat\Psi^\dagger(\xv)\hat\Psi^\dagger(\yv)V(|\xv-\yv|)
     \hat\Psi(\yv)\hat\Psi(\xv),\\
 &H_\mathrm{1B}(\xv)
 = -\frac{\hbar^2\nabla^2_{\xv}}{2m} + V_\mathrm{trap}(\xv),
\label{eq:1BHamiltonian}
\end{align}
The second term on the right hand side of Eq.~(\ref{eq:MBHamiltonian}) represents the bare interaction term.
The bare coupling constants equal the binary interaction  potential $V(|\xv-\yv|)$ at the possible interparticle distances $|\xv-\yv|$.
This potential is obtained, e.g., by describing the quantum mechanics of two-atom interactions in the Born-Oppenheimer approximation, where $|\xv-\yv|$ denotes the internuclear distance.

The dynamics of the non-relativistic many-body system can be described, e.g., in the Schr\"odinger picture, where the many-body state at time $t$ is given  by some density matrix $\rho_D(t)$. 
The Schr\"odinger equation, with the Hamiltonian (\ref{eq:MBHamiltonian}), then completely determines the non-equilibrium dynamics. 
All information about the quantum theory can be encoded in the infinite series of correlation functions or $n$-point functions
\begin{align}
\label{eq:CorrFunc}
  \langle \hat\Psi^\dagger(x_1)\cdots\hat\Psi(x_n)\rangle_t
  &= \Tr[\hat\rho_D(t)\hat\Psi^\dagger(x_1)\cdots\hat\Psi(x_n)],
\end{align}
with $n \ge 1$. For the Hamiltonian (\ref{eq:MBHamiltonian}), the equation of motion of any particular $n$-point function involves other correlation functions up to the order of $n+2$.
Hence, the result is an infinite system of coupled dynamic equations for the correlation functions (cf., e.g., Ref.~\cite{Kohler2002a,Kohler2003a}). 
Since this cannot be solved exactly one has to find suitable approximation schemes. 

Here it is important to note that for out-of-equilibrium calculations there are additional complications which do not appear in vacuum or thermal equilibrium. 
The first new aspect concerns secularity:
Even for weak couplings the strict perturbative time evolution suffers from the presence of spurious, so-called secular, terms which grow with time and invalidate the expansion. 
Moreover, the very same problem appears as well for non-perturbative approximation schemes such as standard $1/{\cal N}$ expansions based on the one-particle irreducible (1PI) effective action~\footnote{Secularity enters the required next-to-leading order corrections of the 1PI $1/{\cal N}$ expansion and beyond~\cite{Mihaila1997a,Ryzhov2000a}.}.
Similar problems can also appear by simply truncating the infinite system of coupled dynamic equations for the correlation functions at a given level of $n$-point functions. 

The problem of secularity has been discussed in Refs. \cite{Berges2004c,Berges2005a}.
Typically, for a given approximation, there can be various ways to resolve the secularity problem by resummation.
There is a requirement, however, which poses very strong restrictions on the possible approximations: 
Universality, i.e.~the insensitivity of the late-time behavior to the details of the initial conditions. 
If thermal equilibrium is approached then the late-time result is universal in the sense that it becomes uniquely determined by the conserved energy density and particle number.
To implement the necessary nonlinear dynamics which recover detailed balance at late times is demanding.
Both requirements of a non-secular and universal behavior can indeed be fulfilled using systematic expansions of the 2PI effective action, which provide a practical means to describe far-from-equilibrium dynamics as well as thermalization from first principles~\cite{Berges2004c,Berges2005a}. 

As will be described below, to lowest order, if all quantum-statistical fluctuations are neglected, the 2PI effective action leads to the Gross-Pitaevskii equation for the mean field 
\footnote{Choosing a local coupling $V(x-y)=g\delta(x-y)$ related to the $s$-wave scattering length, $g=4\pi\hbar^2a/m$, yields an effective theory valid at low energies, cf.~Sect.~\ref{sec:HFB}.}. 
For a sufficiently large mean field this classical field-theory approximation can be used to approximately describe the dynamics of weakly interacting Bose-Einstein condensates. 
Of course, the classical approximation cannot be used to describe the approach to quantum thermal equilibrium characterized by a Bose-Einstein distribution. 

The 2PI effective action, with fluctuations taken into account to two-loop order, describes the (time-dependent) Hartree-Fock-Bogoliubov (HFB) approximation,
in which exchange between the condensate and the non-condensed fraction of the gas is accounted for. 
It conserves total particle number and energy but neglects multiple scattering. 
As a consequence, the approximation fails to describe thermalization. 
Typically, the maximum time for which it is reliable, decreases with increasing interaction strength, e.g., in the case of ultracold alkali atoms, scattering length $a$. 
In particular, the Hartree-Fock approximation is known to suffer from the presence of an infinite series of additional conserved quantities, which are not present in the fully interacting theory. 
These spurious constants of motion are associated to an infinite life-time of quasi-particle momentum modes, which prevent relaxation to a thermal distribution~\cite{Berges2002a}.

The extensively employed HFB equations of motion may also be obtained by rewriting the hierarchy of dynamic equations for the correlation functions  (\ref{eq:CorrFunc}) in terms of their connected counterparts, or cumulants, and neglecting all cumulants of three and more field operators \cite{Kokkelmans2002b,Kohler2002a}.
One finds that the HFB dynamic equation for the leading order cumulant, the mean field, involves a resummation of infinitely many graphs and therefore arbitrary high powers of the bare coupling $V$. 
For Bose-Einstein condensates, many approximation schemes beyond HFB have been introduced, e.g., in Refs.~\cite{Proukakis1996a,Kohler2002a}, and take into account correlation functions of third and higher orders. 
However, the requirements for a reliable late-time behaviour are difficult to implement. 
The approximation schemes, first, have to yield controlled approximations and,
second, must not violate crucial conservation laws like energy conservation,
or, for a non-relativistic gas, the conservation of total particle number.

We emphasize that all these requirements can be met by systematic expansions of the 2PI effective action. 
These include 2PI loop-expansions, coupling-expansions, as well as $1/{\cal N}$-expansions. 
In the following we concentrate on the expansion in inverse powers of $\cN$, where $\cN$ is the number of field components of a scalar theory invariant under the symmetry transformations of the orthogonal group $O(\cN)$. 
Compared to a coupling expansion, this procedure has the advantage that it can be employed in the absence of a weak coupling.
In particular, it can be used to describe the non-analytic dynamics near the  second-order phase transition~\cite{Alford2004a}, where the condensate vanishes --- a situation which cannot be quantitatively described in terms of a coupling expansion.
The scalar theory defined by Eq.~(\ref{eq:MBHamiltonian}) is $U(1)$-invariant,  which is equivalent to an $O(2)$-symmetry. 
In contrast to the standard $1/{\cal N}$ expansion of the 1PI effective action, the apparently rapid convergence of the 2PI expansion even for small values of $\cal N$ is crucial for our approach~\cite{Aarts2002a,Alford2004a}.
In the following we present the 2PI effective action approach to the dynamics of a strongly interacting ultracold Bose-gas.

\section{2PI effective-action approach}
\label{sec:EffActionApproach}
\subsection{Non-equilibrium quantum field theory}
\label{sec:NEqQFT}

Before introducing the 2PI effective action we would like to recall, for the purpose of making our article sufficiently self-contained, some basics about non-equilibrium quantum field theory.
For more details, cf., e.g., Ref. \cite{Berges2005a}.
All information about a non-equilibrium quantum many-body system is contained in the generating functional for non-equilibrium correlation functions %
\footnote{Here and in the following we use natural units where $\hbar=1$. This can be achieved by redefining the mass as $m\to \hbar m$, the external potential as $V_\mathrm{trap}\to\hbar V_\mathrm{trap}$, and the interaction potential as $V\to\hbar V$.
Dividing then the full quantum action $S$ by $\hbar$ leaves the equations of motion, derived in general from the variational principle $\delta S=0$, invariant, but no explicit factors of $\hbar$ appear.}:
\begin{align}
\label{eq:NEgenFuncZ}
  Z[J,K;\hat\rho_D]
  &= \mathrm{Tr}\Big[\hat\rho_D(t_0){\cal T}_{\cal C}
     \exp\Big\{i
     \int_{x}\,\hat\Phi(x)J(x)
     \nonumber\\
  &\qquad
     +\frac{i}{2}\int_{xy}\,\hat\Phi(x)K(x,y)\hat\Phi(y)\Big\}\Big].
\end{align} 
Any correlation function of a quantum many-body system can be derived from this by functional differentiation with respect to $J(x)$ and subsequently setting $J\equiv K\equiv0$. For example, the two-point function follows as
\begin{align}
\label{eq:CorrFuncbyFuncDiff}
  \langle{\cal T}_{\cal C}\hat\Phi(x)\hat\Phi(y)\rangle
  &\equiv\mathrm{Tr}[\hat\rho_D(t_0){\cal T}_{\cal C}\hat\Phi(x)\hat\Phi(y)]
  \nonumber\\
  &= \left.\frac{\delta^2Z[J,K;\hat\rho_D]}
                {i\delta J(x)i\delta J(y)}\right|_{J=K=0}.
\end{align}
Here, $\hat\rho_D(t_0)$ is the normalized density matrix describing the many-body system at the initial time $t_0$, which, in general, does not have the equilibrium form $\rho_D^{(\mathrm{eq})}\sim\exp\{-\beta H\}$. 
$\hat\Phi$ is the operator of a $\cN$-component scalar quantum field, with its space-time arguments $x=(t,\xv)=(x_0,\xv)$.
In the following, where obvious, we suppress indices $i$ enumerating the two components $\hat\Phi_i$, corresponding to the real and imaginary parts of a single complex Bose field $\hat\Psi$, or, to their independent combinations $\hat\Psi$ and $\hat\Psi^*$.

In Eqs.~(\ref{eq:NEgenFuncZ}) and (\ref{eq:CorrFuncbyFuncDiff}), ${\cal T}_{\cal C}$ denotes time ordering along a closed time-path contour $\cal C$ appearing in the source term integrals with $\int_x\equiv\int_{\cal C}dx_0\int d^3x$~\cite{Schwinger1961a,Keldysh1964a}.
The time path $\cal C$ extends from the initial time $t_0$ to some finite time $t>t_0$, and, back, from $t$ to $t_0$.  
Note, that in Eq.~(\ref{eq:CorrFuncbyFuncDiff}), the real-time contour must contain the times of interest, i.e., $x_0$ and $y_0$. In practice, this is no problem since the largest time, ${\rm max}(x_0,y_0)$, is kept as a variable which evolves in the time evolution equations for the correlators dicussed below. 
The second half of $\cal C$, from $t$ to $t_0$, ensures the normalization of the generating functional $Z[0,0;\hat\rho_D]=1$, i.e., unity of the trace of the density matrix.

The generating functional $Z[J,K;\hat\rho_D]$ has a functional integral
representation, which can be found by inserting, to the left and right of $\rho_D$, a complete set of eigenstates of the Heisenberg field operators at the initial time, $\hat\Phi(x_0=t_0,\xv)|\Phi^\pm\rangle=\Phi^\pm(\xv)|\Phi^\pm\rangle$, with $\Phi^\pm(\xv)\equiv\Phi(t_0,\xv)$:
\begin{align}
\label{eq:NEgenFuncZphipm}
  Z[J,K;\hat\rho_D]
  &= \int d\Phi^+d\Phi^-\langle\Phi^+|\hat\rho_D(t_0)|\Phi^-\rangle
  \nonumber\\
  &\quad\times
     \int_{\Phi^+}^{\Phi^-}{\cal D}\Phi
     \exp\Big\{i\Big[S[\Phi]+
     \int_{x}\,\Phi(x)J(x)
     \nonumber\\
  &\qquad
     +\frac{1}{2}\int_{xy}\,\Phi(x)K(x,y)\Phi(y)\Big]\Big\}.
\end{align} 
The functional integral $\int\cD\Phi=\prod_i\int\cD\Phi_i$ sums over all field configurations.
$S$ denotes the classical action, defined as 
\begin{align}
\label{eq:ClassActL}
  S[\Phi]
  &= \int_{x}{\cal L}(x),
\end{align} 
where $\cal L$ is the Lagrangian density.
We will provide $S[\Phi]$ below.

A general initial density matrix can be parametrized as
\begin{align}
\label{eq:IniDMParam}
  \langle\Phi^+|\hat\rho_D(t_0)|\Phi^-\rangle
  &= {\cal N}\exp\{if_{\cal C}[\Phi]\},
\end{align} 
with a normalization $\cN$ and the functional $f$ expanded in powers of the fields:
\begin{align}
\label{eq:fcalC}
  f_{\cal C}[\Phi]
  = \alpha_0 + \sum_{n=1}^\infty\frac{1}{n!}\int_{x_1...x_n}
    \alpha_n(x_1,...,x_n)
    \prod_{i=1}^n\Phi(x_i).
\end{align} 
Here, the coefficients $\alpha_n(x_1,...,x_n)$ vanish identically for all times different from $t_0$  (cf., e.g., Ref.~\cite{Berges2005a}). 

In many practical cases it is sufficient to specify, at time $t_0$, only the lowest correlation functions.
If an initial state is fully determined by the mean field
\begin{align}
\label{eq:MFphi}
  \phi_i(x) 
  &= \langle\hat\Phi_i(x)\rangle
\end{align} 
and the connected two-point function or cumulant
\begin{align}
\label{eq:Gconn}
  G_{ij}(x,y)
  &= \langle\cT_{\cal C}\hat\Phi_i(x)\hat\Phi_j(y)\rangle - \phi_i(x)\phi_j(y)
\end{align} 
then the initial density matrix, Eq.~(\ref{eq:IniDMParam}), can be written as a Gaussian in the field $\Phi$, i.e., all $\alpha_n$ with $n\ge3$ vanish identically.

In this case, comparing Eqs.~(\ref{eq:IniDMParam}) and (\ref{eq:fcalC}) with Eq.~(\ref{eq:NEgenFuncZphipm}), one finds that the initial-time sources can be absorbed into the functional integral by redefining the source fields at times $x_0=y_0=t_0$ according to $J(x)\to J(x)-\alpha_1(x)$ and $K(x,y)\to K(x,y)-\alpha_2(x,y)$.
$\alpha_0$ yields an irrelevant normalization constant.
In this way we arrive at the non-equilibrium generating functional in the form
\begin{align}
\label{eq:genFuncZ}
  Z[J,K] 
  &= \int{\cal D}\Phi\,\exp\Big\{i\Big[
     S[\Phi] + \int_{x}\,\Phi(x)J(x)
     \nonumber\\
  &\qquad
     +\frac{1}{2}\int_{xy}\,\Phi(x)K(x,y)\Phi(y)\Big]\Big\}.
\end{align} 
We emphasize that the use of a Gaussian initial density matrix only restricts  the ``experimental'' setup described by the initial conditions for correlation functions --- higher irreducible correlations can build up corresponding to a non-Gaussian density matrix for times $t > t_0$. 
Non-Gaussian initial density matrices pose no principal problems but require taking into account additional initial-time source fields.

The 2PI effective action is obtained below from a double Legendre transform of the generating functional (\ref{eq:genFuncZ}) with respect to the linear and bilinear source terms, $J(x)$ and $K(x,y)$.
Accordingly, more complicated initial conditions involving higher irreducible correlation functions are most efficiently described in terms of $n$PI effective actions with $n > 2$~\cite{Berges2004a}.
In the following we will always assume that, in order to fully specify the initial state of the many-body system, it is sufficient to provide the initial one- and two-point functions $\phi$ and $G$.

\subsection{2PI effective action}
\label{sec:2PIEA}

The 2PI effective action~\cite{Luttinger1960a,Baym1962a,Cornwall1974a} is defined as a Legendre
transform of the generating functional of connected Greens functions $W[J,K]$, defined by 
\begin{align}
\label{eq:genFuncW}
  Z[J,K] &= \exp\{iW[J,K]\}.
\end{align} 
The double Legendre transform reads
\begin{align}
\label{eq:2PIEA}
  \Gamma[\phi,G]
  &= W[J,K] - \int_{x}\,\phi(x)J(x)
  \nonumber\\
  &\qquad
     -\frac{1}{2}\int_{xy}\,\phi(x)K(x,y)\phi(y)
  \nonumber\\
  &\qquad
     -\frac{1}{2}\int_{xy}\,G(x,y)K(x,y).
\end{align} 
Here, $J$ and $K$ are fixed through the conditions
\begin{align}
\label{eq:2PILegendreCondJ}
  \frac{\delta W[J,K]}{\delta J_i(x)}
  &= \phi_i(x),\\
\label{eq:2PILegendreCondK}
  \frac{\delta W[J,K]}{\delta K_{ij}(x,y)}
  &= \frac{1}{2}\left[\phi_i(x)\phi_j(y)+G_{ij}(x,y)\right],
\end{align} 
with the mean field $\phi_i(x)$ and the two-point function $G_{ij}(x,y)$ defined in Eqs.~(\ref{eq:MFphi}) and (\ref{eq:Gconn}), respectively.

The equations of motion for $\phi_i(x)$ and $G_{ij}(x,y)$ are given by
the stationarity requirements 
\begin{align}
\label{eq:2PIStatCondPhi}
  \frac{\delta\Gamma[\phi,G]}{\delta \phi(x)} &= 0,\\
\label{eq:2PIStatCondG}
  \frac{\delta\Gamma[\phi,G]}{\delta G(x,y)} &= 0,
\end{align} 
which follow directly from the definition (\ref{eq:2PIEA}) for vanishing sources $J$ and $K$.

Eq.~(\ref{eq:2PIEA}) shows that the conventional 1PI effective action $\Gamma[\phi]$ is merely $\Gamma[\phi,G]$ for $K=0$, i.e., it is equivalent to $\Gamma[\phi,G]$ for that function $G$ for which the stationarity condition (\ref{eq:2PIStatCondG}) is fulfilled.
Hence, the equation of motion following from the condition (\ref{eq:2PIStatCondPhi}) is equivalent to that derived from the 1PI effective action. 
On the exact level all effective actions are, of course, equivalent. 
Here the particular reason for using the 2PI effective action is that it efficiently allows to devise systematic approximation schemes suitable for nonequilibrium dynamics.

It is convenient to write the 2PI effective action as 
\begin{align}
\label{eq:2PIEAexp}
  \Gamma[\phi,G]
  &= S[\phi] +\frac{i}{2}\,\Tr\left\{\ln G^{-1}+G_0^{-1}[\phi]G\right\} 
     +\Gamma_2[\phi,G]\nonumber\\
  &\qquad+\mathrm{const.}\,,
\end{align} 
which contains the contribution from the classical action $S$, a one-loop-type
term and a term $\Gamma_2[\phi,G]$ that contains all the rest.
The trace, the logarithm and the product of Greens functions in the third term are meant in the functional sense.
$G_0^{-1}$ is the inverse of the classical propagator
\begin{align}
\label{eq:G0inv}
  iG_{0,ij}^{-1}(x,y;\phi)
  &= \frac{\delta^2S[\phi]}{\delta\phi_i(x)\delta\phi_j(y)},
\end{align} 
for which we will give an explicit expression in the next section.

Taking the derivative of (\ref{eq:2PIEAexp}) in terms of $G$ one observes that the second stationarity condition (\ref{eq:2PIStatCondG}) is equivalent to the exact Dyson-Schwinger equation for the propagator,
\begin{align}
\label{eq:DysonSchwinger}
  G_{ij}^{-1}(x,y)
  &= G_{0,ij}^{-1}(x,y;\phi)-\Sigma_{ij}(x,y;\phi,G),
\end{align} 
with the proper self-energy
\begin{align}
\label{eq:Sigma}
  \Sigma_{ij}(x,y;\phi,G)
  &= {2i}\frac{\delta\Gamma_2[\phi,G]}{\delta G_{ij}(x,y)}.
\end{align} 
From Eq.~(\ref{eq:DysonSchwinger}) one easily sees that the proper self-energy $\Sigma$ is 1PI.
Since the functional differentiation of $\Gamma_2$ with respect to $G$ corresponds to opening one propagator line in any diagram contributing to $\Gamma_2$, the expansion of $\Gamma_2$ may only contain diagrams which are at least two-particle irreducible.
This feature constitutes the name of the 2PI effective action.

We note that the above mentioned, more general $n$PI effective actions can be constructed analogously.
These depend on correlation functions up to order $n$ and generate diagrammatic expansions in which also the corresponding higher vertices are self-consistently determined, i.e.~dressed~\cite{Berges2005a}. 

Before we apply the effective-action description of a non-equilibrium many-body system to derive dynamic equations we close this section by introducing the spectral,
\begin{align}
\label{eq:rhoij}
  \rho_{ij}(x,y) 
  &= i\langle [\hat{\Phi}_i(x),\hat{\Phi}_j(y)] \rangle_c,
\end{align} 
and statistical components,
\begin{align}
\label{eq:Fij}
  F_{ij}(x,y) 
  &= \frac{1}{2}\langle \{\hat{\Phi}_i(x),\hat{\Phi}_j(y)\}\rangle_c,
\end{align} 
of the two-point function $G$, where the subscript $c$ means that, for $\phi\not=0$, disconnected parts are substracted. 
In Eq.~(\ref{eq:Fij}), $\{,\}$ denotes the anticommutator.
The decomposition identity for the two-point function $G$ follows from Eqs.~(\ref{eq:Gconn}), (\ref{eq:rhoij}), and (\ref{eq:Fij}) as 
\begin{align}
\label{eq:GintoFandrho}
  G_{ij}(x,y)
  &= F_{ij}(x,y) -\frac{i}{2}\rho_{ij}(x,y)\,\mathrm{sign}_{\cal C}(x_0-y_0),
\end{align} 
where $\mathrm{sign}_{\cal C}(x_0-y_0)$ is the sign function along the path
$\cal C$ which evaluates to $1$ ($-1$) if $x_0$ ($y_0$) is prior to 
$y_0$ ($x_0$) along $\cal C$.

While the spectral function encodes the spectrum of the theory, the statistical propagator gives information about occupation numbers.
Loosely speaking, the decomposition makes explicit what states are available and how often they are occupied.
Note that, in thermal equilibrium, the spectral and statistical functions would be related by the fluctuation-dissipation relation, which, for a non-condensed homogeneous system in energy-momentum space, reads
\begin{align}
\label{eq:FlucDiss}
  F^\mathrm{(eq)}(\omega,\pv)
  &= -i\left(\frac{1}{2}+n(\omega,T)\right)\rho^\mathrm{(eq)}(\omega,\pv),
\end{align} 
with the Bose-Einstein distribution function
$n(\omega,T)=(e^{(\omega-\mu)/k_\mathrm{B}T}-1)^{-1}$.
Relation (\ref{eq:FlucDiss}) does no longer apply in a system far from equilibrium, such that the dynamical evolution of $F$ and $\rho$ is given by separate (coupled) equations of motion.

\subsection{Dynamic equations}
\label{sec:DynEq}
In this section, we derive dynamic equations for the mean field $\phi$ and the two-point cumulant $G$ from the stationarity conditions, Eqs.~(\ref{eq:2PIStatCondPhi}) and (\ref{eq:2PIStatCondG}).
We consider explicitly the theory defined by the Hamiltonian (\ref{eq:MBHamiltonian}) and choose the representation
\begin{align}
\label{eq:Phi1Phi2}
  &\Phi_1(x)=\sqrt{2}\mathrm{Re}\,\Psi(x),\qquad
   \Phi_2(x)=\sqrt{2}\mathrm{Im}\,\Psi(x)
\end{align} 
of the quantum field in terms of its real and imaginary parts.
The corresponding field operators obey the bosonic commutation relation $[\hat\Phi_1(\xv,t),\hat\Phi_2(\yv,t)]=i\delta(\xv-\yv)$ while all other equal-time commutators vanish.
This representation will be particularly convenient when discussing the non-perturbative $1/\cN$ expansion in Chapter \ref{sec:Nonpert}.
The classical action is given by
\begin{align}
\label{eq:ClassAct}
  S[\Phi]
  &= \frac{1}{2}\int_{xy}\,
     \Phi_i(x)\,iD^{-1}_{ij}(x-y)\Phi_j(y)+S_\mathrm{int}[\Phi],
\end{align} 
with the inverse free propagator (Greens function)
\begin{align}
\label{eq:invfreeProp}
  iD^{-1}_{ij}(x-y)
  &= \delta_{\cal C}(x-y)\left(-i\sigma_{2,ij}\ptd_{y_0}
   -H_\mathrm{1B}(\xv)\delta_{ij}
     \right),
\end{align} 
containing the Pauli matrix
\begin{align}
\label{eq:PauliMat2}
  \sigma_2 = \left(\ba{cc}0&-i\\i&0\ea\right).
\end{align} 
Here, $\delta_{\cal C}(x-y)\equiv\delta_{\cal C}(x_0-y_0)\delta^{(3)}(\xv-\yv)$ denotes the four-dimensional Dirac distribution on the closed time path.
The interaction part $S_\mathrm{int}[\Phi]$ may be expressed, for the Hamiltonian (\ref{eq:MBHamiltonian}), with
$V(x-y)\equiv V(|\xv-\yv|,x_0)\delta_{x_0,y_0}$, as
\begin{align}
\label{eq:Sint}
  S_\mathrm{int}[\Phi]
  &= -\frac{1}{8}\int_{xy}\,
     \Phi_i(x)\Phi_j(y)V(x-y)\Phi_j(y)\Phi_i(x).
\end{align} 
where it is summed over double indices.
We use Eqs.~(\ref{eq:ClassAct}), (\ref{eq:Sint}) to derive the classical inverse propagator, cf.~Eq.~(\ref{eq:G0inv}):
\begin{align}
\label{eq:G0invexpl}
  &iG_{0,ij}^{-1}(x,y;\phi)
  = \frac{\delta^2S[\phi]}{\delta\phi_i(x)\delta\phi_j(y)}
  \nonumber\\
  &\quad= iD^{-1}_{ij}(x-y)
     -\phi_i(x)V(x-y)\,\phi_j(y)
    \nonumber\\
  &\qquad-\frac{1}{2}\int_{z}\,\phi_k(z)V(x-z)\,\phi_k(z)
      \delta_{ij}\delta_{\cal C}(x-y),
\end{align} 
Now, everything is prepared to derive the dynamic equations by imposing the stationarity conditions, Eqs.~(\ref{eq:2PIStatCondPhi}) and (\ref{eq:2PIStatCondG}).
These conditions yield, together with Eqs.~(\ref{eq:2PIEAexp}), (\ref{eq:ClassAct}), (\ref{eq:Sint}), and (\ref{eq:G0invexpl}):
\begin{align}
\label{eq:2PIStatCondPhiexpl}
  0&=\delta\Gamma[\phi,G]/\delta\phi_i(x)
  \nonumber\\
  &=\int_{y}\,\Big[
    iD^{-1}_{ij}(x-y)\,\phi_j(x)
    \nonumber\\
  &\qquad
    -V(x-y)\frac{1}{2}\Big(\phi_k(y)\phi_k(y)+G_{kk}(y,y)\Big)
    \phi_i(x)
    \nonumber\\
  &\qquad
    -V(x-y)\,G_{ij}(x,y)
    \,\phi_j(y)\Big]
    \nonumber\\
  &\quad
    +\delta\Gamma_2[\phi,G]/\delta\phi_i(x),
  \\[0.2cm]
\label{eq:2PIStatCondGexpl}
  0&=\delta\Gamma[\phi,G]/\delta G_{ij}(x,y)
  \nonumber\\
  &=\frac{i}{2}\left[G_{0,ji}^{-1}(y,x,\phi)-G^{-1}_{ji}(y,x)
    \right]
    \nonumber\\
  &\quad
    +\delta\Gamma_2[\phi,G]/\delta G_{ij}(x,y).
\end{align} 
From Eq.~(\ref{eq:2PIStatCondPhiexpl}), the equation of motion for the mean field $\phi_i$ is obtained as
\begin{align}
\label{eq:EOMphi}
   &\big[-i\sigma_{2,ij}\partial_{x_0}-H_\mathrm{1B}(\xv)\delta_{ij}\big]\,
    \phi_j(x)
   \nonumber\\
   &\ = \int_{y}\,V(x-y)\Big[
     \frac{1}{2}\Big(\phi_k(y)\phi_k(y)+G_{kk}(y,y)\Big)\phi_i(x)
   \nonumber\\
   &\qquad
    +G_{ij}(x,y)\phi_j(y)\Big]
   \nonumber\\
   &\quad
   -\delta\Gamma_2[\phi,G]/\delta\phi_i(x).
\end{align} 
The dynamic equation for the Greens function $G$ follows from
Eq.~(\ref{eq:2PIStatCondGexpl}) by convolution 
with $G$ and using the definition (\ref{eq:Sigma}) of the self-energy $\Sigma$:
\begin{align}
\label{eq:EOMG}
  &\int_{z}\ G^{-1}_{0,ik}(x,z;\phi)G_{kj}(z,y)
  \nonumber\\
  &\quad=\ \delta_{\cal C}(x-y)\delta_{ij}
   +\int_{z}\ \Sigma_{ik}(x,z;\phi,G)G_{kj}(z,y).
\end{align} 
Once the 2PI part $\Gamma_2[\phi,G]$ of the effective action is known, Eqs.~(\ref{eq:EOMphi}) and (\ref{eq:EOMG}), representing a closed system of dynamic equations for $\phi$ and $G$, may be solved to determine, for given initial correlation functions $\phi(t_0)$ and $G(t_0,t_0)$, the complete dynamical evolution of the system. 
It is, of course, not possible to solve the equations exactly.
In the following we discuss approximation
schemes for $\Gamma_2$ and use these to obtain approximative dynamic equations for $\phi$ and $G$.

\subsection{Number conservation} 
\label{sec:NumCons}
A major advantage of the effective action approach is that it automatically provides us with dynamic equations which are particle number conserving.
This is a consequence of the Noether theorem in conjunction with the invariance of the theory under orthogonal transformations and can be seen as follows %
\footnote{Conservation of total particle number in a non-relativistic system corresponds to conservation of the difference of particles and antiparticles in a fully relativistic approach, i.e., to the conservation of charge. Neglecting the antiparticle sector of the Hilbert space in a non-relativistic system is equivalent to the constraint of a vanishing antiparticle number and, hence, amounts to the fact that the $O(2)$ symmetry of the Lagrangian ensures total number conservation.}:
The stationarity conditions (\ref{eq:2PIStatCondPhi}) and (\ref{eq:2PIStatCondG}) for the mean field and the propagator can be combined to the equation
\begin{align}
\label{eq:StatCondCombiforNC}
  \sigma_{2,ij}\Big[
  \phi_i(x)\frac{\delta\Gamma[\phi,G]}{\delta\phi_j(x)}
  +2\int_y\frac{\delta\Gamma[\phi,G]}{\delta G_{kj}(y,x)}G_{ki}(y,x)\Big]
  &=0,
\end{align} 
with the elements $\sigma_{2,ij}$ of the Pauli matrix (\ref{eq:PauliMat2}).
From the specific expression (\ref{eq:2PIEAexp}) for the 2PI effective action follows that Eq.~(\ref{eq:StatCondCombiforNC}) is equivalent to the relation
\begin{align}
\label{eq:ContEq}
  &\partial_{x_0} n(x)-\nabv\vec{j}(x)
  \nonumber\\
  & = -{2i}\sigma_{2,ij}\Big[
  \phi_i(x)\frac{\delta\Gamma_\mathrm{int}[\phi,G]}
                {\delta\phi_j(x)}
  +2\int_y \frac{\delta\Gamma_\mathrm{int}[\phi,G]}
                {\delta G_{kj}(y,x)}G_{ki}(y,x)
  \Big].
\end{align} 
Here,
\begin{align}
\label{eq:totaldensity}
  n(x)
  &= \phi_i(x)\phi_i(x)+G_{ii}(x,x),
  \\
\label{eq:totalcurrentdensity}
  \vec{j}(x)
  &= \frac{1}{m}\big[\phi_2(x)\nabv\phi_1(x)-\phi_1(x)\nabv\phi_2(x)
  \nonumber\\
  &\qquad+\langle\cT_{\cal C}(
   \hat\Phi_2(x)\nabv\hat\Phi_1(x)-\hat\Phi_1(x)\nabv\hat\Phi_2(x)
   )\rangle_c\big]
\end{align} 
are the total number and current densities, respectively.
Clearly, particle number is conserved locally if $n$ and $\vec{j}$ obey a continuity equation, i.e., if the right hand side of Eq.~(\ref{eq:ContEq}) vanishes identically.
We consider the specific structure of these terms:
The interaction part of the 2PI effective action occuring therein is defined as
\begin{align}
\label{eq:Gamma2PIint}
   \Gamma_\mathrm{int}[\phi,G]
   = \Gamma[\phi,G] 
   &- \frac{1}{2}\int_{xy}\,
      \phi_i(x)\,iD^{-1}_{ij}(x-y)\phi_j(y)
   \nonumber\\
   &-\frac{i}{2}\mathrm{Tr}\left[D^{-1}G\right].
\end{align} 
The 2PI effective action is, like the underlying quantum action $S[\Phi]$, Eqs.~(\ref{eq:ClassAct}) and (\ref{eq:Sint}), a singlet under $O(2)$ rotations.
It is parametrized by the classical fields $\phi_i$ and $G_{ij}$, where the number of $\phi$-fields has to be even in order to construct an $O(2)$-singlet. 
From the fields $\phi_i$ alone one can construct only one independent invariant under $O(2)$ rotations, which can be taken as $\tr(\phi\phi) \equiv \phi^2 = \phi_i \phi_i$. All functions of $\phi$ and $G$, which are singlets under $O(2)$, can be built from the irreducible, i.e., in field-index space not factorizable, invariants~\cite{Berges2002a,Aarts2002b}
\begin{align}
\label{eq:ONinvariants}
  \phi^2, 
  \quad\quad 
  \tr (G^n), 
  \quad\quad \mbox{and}  \quad\quad 
  \tr (\phi \phi G^{n}), 
\end{align}
with $n=1,2,...$.
As before, the trace tr$(\cdot)$ only applies to the field-component indices while there is no integration over space-time, e.g., tr$(G^3)\equiv G_{ij}(x,y)G_{jk}(y,z)G_{ki}(z,x)$.

For contributions to $\Gamma_\mathrm{int}$ which contain only $\phi^2$ or tr$(G^n)$, the terms in square brackets in Eq.~(\ref{eq:ContEq}) either vanish separately or are symmetric under the exchange of $i$ and $j$.
Moreover, if a term contains an invariant of the form tr$(\phi\phi G^n)$, as, e.g., the contributions remaining in $\Gamma_\mathrm{int}$ from Tr$\{G_0^{-1}G\}$, the combination of the terms in square brackets in Eq.~(\ref{eq:ContEq}) is symmetric under transposition in field index space.
Hence, the total number density is conserved locally as a consequence of the $O(2)$ symmetry of the theory, and, more importantly, this is true for any set of approximative dynamic equations derived from a truncated but still $O(2)$-symmetric effective action.
Note, finally, that only terms in the action which contain mixed invariants tr$(\phi\phi G^n)$ induce exchange of particles between the condensate and the non-condensed fraction of the gas.

\section{Diagrammatic representation} 
\label{sec:PT}
%
\subsection{Loop expansion of $\Gamma_2[\phi,G]$} 
\label{sec:PertGamma2}
In this section we briefly discuss the diagrammatic methods that we will use later on to generate the equations of motion for $\phi$ and $G$. 
For a more comprehensive review see, e.g., Ref.~\cite{Berges2005a} and references cited therein.

\begin{figure}[tb]
\begin{center}
\includegraphics[width=0.47\textwidth]{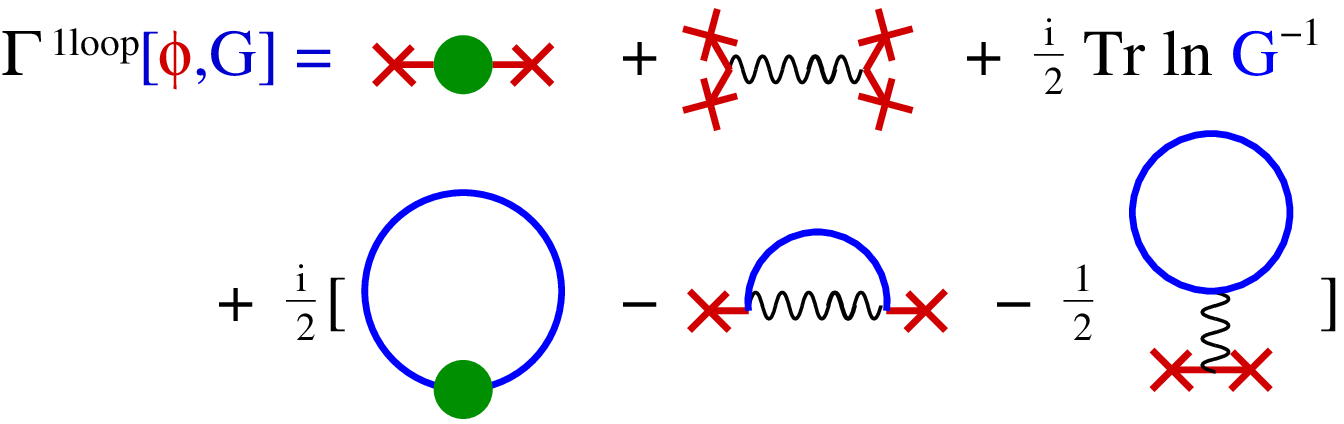}
\end{center}
\vspace*{-3ex}
\caption{
(Color online) Diagrammatic representation of the one-loop part $\Gamma^\mathrm{1loop}[\phi,G]=\Gamma[\phi,G]-\Gamma_2[\phi,G]$ of the 2PI effective action, cf.~Eq.~(\ref{eq:2PIEAexp}), without the irrelevant constant.
The inverse free propagator $D^{-1}(x,y)$ is represented as a (green) filled circle, external fields $\phi(x)$ as (red) crosses, the bare potential $V(x-y)$ as a wiggly line, and the full propagator $G(x,y)$ as a (blue) solid line -- colors are seen in the online version.
At each `vertex' it is summed over double field indices and integrated over the respective space-time variable $x$.
The first two diagrams are the classical action $S[\phi]$, the last three diagrams the term $\frac{i}{2}\mathrm{Tr}\{G_0^{-1}G\}$, cf.~Eq.~(\ref{eq:G0invexpl}).
}
\label{fig:Gamma2PI1loop}
\end{figure}
The 2PI effective action, Eq.~(\ref{eq:2PIEAexp}), can be expanded into a series of closed loop diagrams, i.e., diagrams without external lines.
In Fig.~\ref{fig:Gamma2PI1loop}, the diagrammatic representation of $\Gamma[\phi,G]-\Gamma_2[\phi,G]$ is shown, cf.~Eq.~(\ref{eq:2PIEAexp}).
\begin{figure}[tb]
\begin{center}
\includegraphics[width=0.45\textwidth]{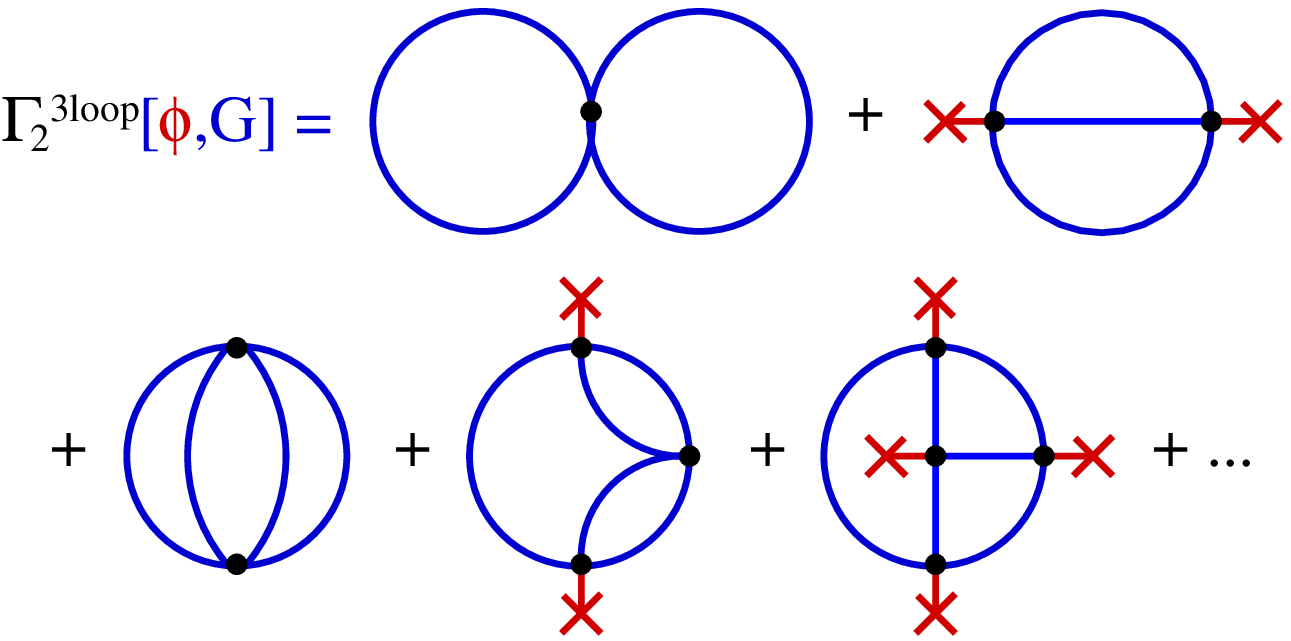}
\end{center}
\vspace*{-3ex}
\caption{
(Color online) Diagrammatic representation of the two- and three-loop diagrams contributing to the 2PI part $\Gamma_2[\phi,G]$ of the 2PI effective action, cf.~Eq.~(\ref{eq:2PIEAexp}).
To simplify the diagrams, the bare vertices $V(x-y)$ are drawn as black dots.
Each such vertex is understood to represent a sum of the topologically different terms shown in Fig.~\protect\ref{fig:nonlocalbareVertex}.
At each vertex, it is summed over double field indices and double space-time variables according to the respective diagram in Fig.~\protect\ref{fig:nonlocalbareVertex}.
}
\label{fig:Gamma23loop}
\end{figure}
Fig.~\ref{fig:Gamma23loop} shows the lowest order diagrams contributing to $\Gamma_2[\phi,G]$.
\begin{figure}[tb]
\begin{center}
\includegraphics[width=0.43\textwidth]{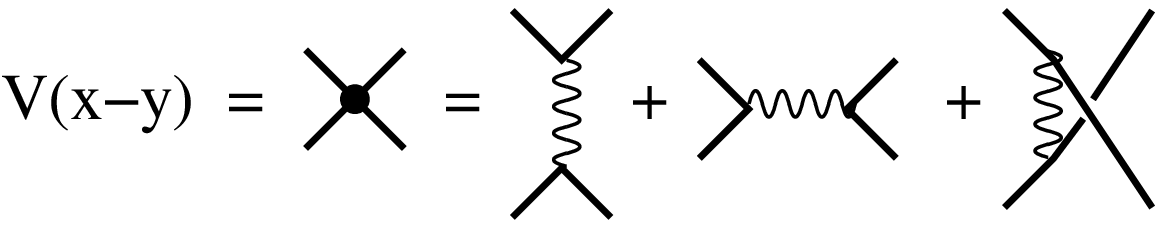}
\end{center}
\vspace*{-3ex}
\caption{
The representation of the bare vertex in terms of a black dot stands for a sum of the three topologically different connections of the four `corners'.
The black lines do not represent propagators and are only drawn in order to illustrate the different possible connections of propagators and/or external fields at the vertices in Figs.~\protect\ref{fig:Gamma2HFBnonloc} and \protect\ref{fig:Gamma2NLO1Nnonloc}
}
\label{fig:nonlocalbareVertex}
\end{figure}
Internal lines in these diagrams represent the full two-point Greens function $G$, vertices the bare interaction $V$.
In addition, one needs to distinguish topologically different diagrams.
This becomes obvious when considering an interaction $V\equiv V(\xv-\yv,x_0)$ which is non-local in its spatial arguments and its diagonal coupling to the field components $1$ and $2$ as exhibited by Eq.~(\ref{eq:Sint}).
An example of this are the diagrams shown in Fig.~\ref{fig:Gamma2HFBnonloc} which combine to give the ``double-bubble'' diagram, i.e., the first graph in the loop expansion of $\Gamma_2$ in Fig.~\ref{fig:Gamma23loop}.

Counting the number of loops, $\Gamma[\phi,G]-\Gamma_2[\phi,G]$ is of leading order, and, as discussed in Appendix \ref{app:Gamma2PI1loop}, this part of the effective action equals the 1PI effective action to one-loop order.
We have seen in Section \ref{sec:2PIEA} that $\Gamma_2$ can only contain 2PI diagrams. 
For this reason it is at least of two-loop order, the leading order comprising the double-bubble and the ``setting-sun'' diagrams in Fig.~\ref{fig:Gamma23loop}.

In the following section we will consider the most simple approximation of $\Gamma_2$ which only takes into account the double-bubble diagram.
The dynamic equations resulting in this approximation constitute the well known time-dependent Hartree-Fock (vanishing mean field, $\phi=0$) or Hartree-Fock-Bogoliubov ($\phi\not=0$) equations which describe coupling between the condensate and the thermal fractions of the gas and account for the formation of pair correlations through binary interactions.

\subsection{Hartree-Fock-Bogoliubov dynamics} 
\label{sec:HFB}
\begin{figure}[tb]
\begin{center}
\includegraphics[width=0.45\textwidth]{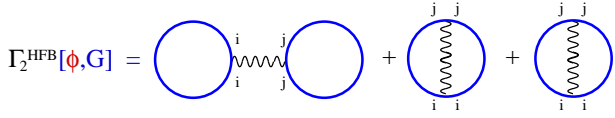}
\end{center}
\vspace*{-3ex}
\caption{
(Color online) Diagrammatic representation of the diagrams contributing, in the Hartree-Fock(-Bogoliubov) approximation, to the 2PI part $\Gamma_2[\phi,G]$ of the 2PI effective action.
At each vertex, it is summed over double field indices which are indicated, and integrated over double space-time variables.
For an explanation of the symbols see the caption of Fig.~\protect\ref{fig:Gamma2PI1loop}.
}
\label{fig:Gamma2HFBnonloc}
\end{figure}

In this section we derive the coupled Hartree-Fock-Bogoliubov equations of motion for the mean field $\phi$ and the correlation function $G$ by evaluating Eqs.~(\ref{eq:EOMphi}) and (\ref{eq:EOMG}) for the leading-order approximation to the 2PI part of the effective action, the double-bubble diagram in Fig.~\ref{fig:Gamma23loop} which is shown, for a non-local coupling $V(x-y)$, in Fig.~\ref{fig:Gamma2HFBnonloc}.

For a better comparison with standard results, we shall, in this section, use the representation where the field $\Phi$ and its complex conjugate are considered as independent quantities.
To this end we write the effective action in terms of a two-component field with
\begin{align}
\label{eq:Phi1PhiPhi2Phistar}
  &\Phi_1(x)=\Psi(x),\qquad\Phi_2(x)=\Psi^*(x),
\end{align} 
Note that the representation in terms of the components defined in (\ref{eq:Phi1PhiPhi2Phistar}) is obtained from the `cartesian' representation (\ref{eq:Phi1Phi2}) by the $U(2)$ unitary transformation
\begin{align}
\label{eq:U2Transform}
  \left(\ba{c}\Phi_1\\ \Phi_2\ea\right)
  \to \frac{1}{\sqrt{2}}\left(\ba{cc}1&i\\1&-i\ea\right)
  \left(\ba{c}\Phi_1\\ \Phi_2\ea\right).
\end{align} 
Hence, introducing an upper-index notation according to
\begin{align}
\label{eq:IndexRaising}
  \Phi^i = \Phi_i^*=\Phi_{3-i},
  \nonumber\\
  G_i^{\ \ 3-j} = G_{ij},\quad \mathrm{etc.,}
\end{align} 
with $i,j\in\{1,2\}$, the double-bubble contribution to $\Gamma_2$ can be written in the form 
\begin{align}
\label{eq:Gamma2HFB}
  \Gamma_2^\mathrm{HFB}[\phi,G]
  &= -\frac{1}{8}\int_{xy}V(x-y)\Big[
     G_i^{\ i}(x,x)G_j^{\ j}(y,y)
  \nonumber\\
  &\phantom{-\frac{1}{8}\int_{xy}V(x-y)}
  +2G_i^{\ j}(x,y)G_j^{\ i}(y,x)\Big],
\end{align} 
where it is again summed over double indices.
The statistical factor $2$ in the second term in brackets accounts for the topologically equal diagrams, i.e., for the second and third diagram in Fig.~\ref{fig:Gamma2HFBnonloc}.

The proper self energy corresponding to the $\Gamma_2^\mathrm{HFB}$ in Eq.~(\ref{eq:Gamma2HFB}) is obtained using Eq.~(\ref{eq:Sigma}):
\begin{align}
\label{eq:SigmaHFB}
   \Sigma_i^{\ j}(x,y)
   &= 2i\frac{\delta\Gamma_2[\phi,G]}{\delta G_i^{\ j}(x,y)}
   \nonumber\\
   &= -\frac{i}{2}\Big\{\int_z V(z-x)G_k^{\ k}(z,z)\delta_i^{\ j}\delta_{\cal C}(x-y)
   \nonumber\\
   &\qquad\quad+2V(x-y) G_j^{\ i}(y,x)\Big\}.
\end{align} 

By making use of the symmetry relations
\begin{align}
\label{eq:SymmF}
   &F_i^{\ j}(x,y) = F^j_{\ i}(y,x) = [F_j^{\ i}(y,x)]^*,
   \\
\label{eq:Symmrho}
   &\rho_i^{\ j}(x,y) = -\rho^j_{\ i}(y,x) = [-\rho_j^{\ i}(y,x)]^*,
\end{align} 
the dynamic equation (\ref{eq:EOMphi}) for the mean field may be reexpressed in terms of $\phi=\phi_1=\phi^2$ and $F$ as
\begin{align}
\label{eq:NLSEHFB}
   &\big[i\hbar\partial_{t}-H_\mathrm{1B}(\xv)\big]\,\phi(\xv,t)
   \nonumber\\
   &\ = \int_{\yv}\,V(\xv-\yv,t)\Big[
     \Big(\phi(\yv,t)\phi(\xv,t)+F_m(\yv,\xv;t)\Big)\phi^*(\yv,t)
   \nonumber\\
   &\qquad\quad
    +F_n(\yv,\xv;t)\,\phi(\yv,t)
    +F_n(\yv,\yv;t)\,\phi(\xv,t)
    \Big].
\end{align} 
In deriving Eq.~(\ref{eq:NLSEHFB}) we have integrated $y_0$ over the path $\cal C$, leaving only the space integral $\int_{\yv}=\int d^3y$, and used the locality of $V(x-y)=V(\xv-\yv,t)\delta_{\cal C}(x_0-y_0)$ in its time argument $t\equiv x_0$.
By this integration, the spectral functions are evaluated at equal times and disappear from the equation, either because of a vanishing equal-time commutator, or through cancellation between different terms.
The statistical two-point functions are also evaluated at equal times, and we denote the different matrix elements occuring as 
\begin{align}
\label{eq:Fn}
  F_n(\xv,\yv;t) 
  &= \left.F_2^{\ 2}(x,y)\right|_{x_0=y_0}
   = \frac{1}{2}\langle\{\hat{\Psi}^\dagger(\xv,t),\hat{\Psi}(\yv,t)\}\rangle_c,
  \\
\label{eq:Fm}
  F_m(\xv,\yv;t) 
  &= \left.F_1^{\ 2}(x,y)\right|_{x_0=y_0}
   = \frac{1}{2}\langle\{\hat{\Psi}(\xv,t),\hat{\Psi}(\xv,t)\}\rangle_c.
\end{align} 
Eq.~(\ref{eq:NLSEHFB}) is the well-known non-linear dynamic equation for the condensate mean field, where the contribution from the three-point connected Greens function or cumulant $\langle\hat{\Psi}^\dagger(\yv)\hat{\Psi}(\yv)\hat{\Psi}(\xv)\rangle_c$ has been neglected, cf., e.g., Ref.~\cite{Kohler2002a}.
The equation, however, involves the symmetrized two-point cumulants instead of the normal ordered ones, the time-dependent non-condensate and anomalous density matrices, $\tilde{n}$ and $\tilde{m}$, respectively:
\begin{align}
\label{eq:ncdmatrix}
  \widetilde{n}(\xv,\yv;t) 
  &= \langle\hat{\Psi}^\dagger(\yv,t)\hat{\Psi}(\xv,t)\rangle_c,
  \\
\label{eq:admatrix}
  \widetilde{m}(\xv,\yv;t) 
  &= \langle\hat{\Psi}(\yv,t)\hat{\Psi}(\xv,t)\rangle_c.
\end{align} 
The reason for this is that the functional integral $Z$, as defined in Eqs.~(\ref{eq:genFuncZ}), (\ref{eq:ClassAct}), is equivalent to a Hilbert-space formulation with a Weyl-ordered Hamiltonian operator $\hat{H}$, i.e., an $\hat{H}$ which is invariant under permutations of non-commuting field operators.
Hence, the effective-action approach yields dynamic equations involving symmetrized correlation functions $F_n$ and $F_m$.
In order to rewrite Eq.~(\ref{eq:NLSEHFB}) in terms of normal ordered cumulants one can use the commutation relations which give $F_n(\xv,\yv;t)=\widetilde{n}(\xv,\yv;t)+\delta(\xv-\yv)/2$ and $F_m(\xv,\yv;t)=\widetilde{m}(\xv,\yv;t)$.
This introduces, in Eq.~(\ref{eq:NLSEHFB}), infinite terms involving the combination $V(0,t)+\int_{\yv}V(\yv,t)\delta(0)$, which also appear if the mean-field equation is derived by means of the Schr\"odinger equation for the density matrix in Eq.~(\ref{eq:CorrFunc}), with a Weyl-ordered Hamiltonian.
These contributions disappear when shifting the zero of the energy scale through normal ordering of the Hamiltonian.

The mean-field equation (\ref{eq:NLSEHFB}) shows that, in order to obtain a closed set of dynamic equations, the equations of motion for $F_n(x,y)$ and $F_m(x,y)$ are required only for equal times, $x_0=y_0$.
The equation for $F_n(x,y)$ is obtained from the equivalent of Eq.~(\ref{eq:EOMG}) in the representation (\ref{eq:Phi1PhiPhi2Phistar}), by substracting the expression for the time derivative $\partial_{y_0}G_2^{\ 1}(y,x)$ from that for $\partial_{x_0}G_1^{\ 2}(x,y)$.
Taking the equal-time limit 
\footnote{The time ordering of $x_0<y_0$ along the path $\cal C$ needs to be fixed in order to obtain an explicit set of dynamic equations.
A different ordering would describe the same physics but would yield, in general, different equations.} 
$x_0=y_0-0$, the resulting equation reads
\begin{align}
\label{eq:EOMFnHFB}
  &\big[i\hbar\partial_{t}-H_\mathrm{1B}(\xv)+H_\mathrm{1B}(\yv)\big]\,
   F_n(\xv,\yv;t)
  \nonumber\\
  &\ =\ \Big\{\int_{\zv}V(\xv-\zv,t)
  \nonumber\\
  &\qquad
   \times\Big[\Big(\phi(\xv,t)\phi(\zv,t)+F_m(\xv,\zv;t)\Big)F_m^*(\zv,\yv;t)
  \nonumber\\
  &\qquad
   +\phi^*(\zv,t)
   \Big(\phi(\zv,t) F_n(\xv,\yv;t)+\phi(\xv,t) F_n(\zv,\yv;t)\Big)
  \nonumber\\
  &\qquad
   +F_n(\zv,\zv;t)F_n(\xv,\yv;t)+F_n(\xv,\zv;t)F_n(\zv,\yv;t)
  \Big]\Big\}
  \nonumber\\
  &\quad-\{\xv\leftrightarrow\yv\}^*.
\end{align} 
The last term denotes the complex conjugate of the first term in curly brackets, with $\xv$ and $\yv$ interchanged.
Similarly, the time evolution of $F_m(\xv,\yv;t)$ is derived by adding the equations for $\partial_{x_0}G_1^{\ 2}(x,y)$ and for $\partial_{y_0}(G_2^{\ 1}(y,x))^*$, and taking the same equal time limit as above:  
\begin{align}
\label{eq:EOMFmHFB}
  &\big[i\hbar\partial_{t}-H_\mathrm{1B}(\xv)-H_\mathrm{1B}(\yv)\big]\,
   F_m(\xv,\yv;t)
  \nonumber\\
  &\ =\ \Big\{\int_{\zv}V(\xv-\zv,t)
  \nonumber\\
  &\qquad
   \times\Big[\Big(\phi(\xv,t)\phi(\zv,t)+F_m(\xv,\zv;t)\Big)F_n(\yv,\zv;t)
  \nonumber\\
  &\qquad
   +\phi^*(\zv,t)
   \Big(\phi(\zv,t) F_m(\xv,\yv;t)+\phi(\xv,t) F_m(\zv,\yv;t)\Big)  \nonumber\\
  &\qquad
   +F_n(\zv,\zv;t)F_m(\xv,\yv;t)+F_n(\xv,\zv;t)F_m(\zv,\yv;t)
  \Big]\Big\}
  \nonumber\\
  &\quad+\{\xv\leftrightarrow\yv\}.
\end{align} 
Reexpressing Eqs.~(\ref{eq:EOMFnHFB}) and (\ref{eq:EOMFmHFB}) in terms of normal ordered cumulants $\widetilde n$ and $\widetilde m$ 
\footnote{In Refs.~\cite{Kohler2002a,Kohler2003a,Goral2004a,Gasenzer2004b}, the following notation was used: $\Psi(\xv;t)=\phi(\xv,t)$, $\Gamma(\xv,\yv;t)=\widetilde{n}(\xv,\yv;t)$, and $\Phi(\xv,\yv;t)=\widetilde{m}(\xv,\yv;t)$.}, 
one recovers, together with the non-linear mean-field equation (\ref{eq:NLSEHFB}), the familiar closed set of dynamical Hartree-Fock-Bogoliubov equations describing the exchange of atoms between the condensate and thermal fractions of the gas; cf., e.g., Ref.~\cite{Kohler2002a}.
These equations have been extensively used to describe the short-time, far-from-equilibrium dynamics of strongly interacting Bose-Einstein condensates.
Applications include the production of correlated atom pairs in four-wave mixing of condensates \cite{Kohler2002a}, 
the atom-molecule coherence in a Ramsey-type interfero\-meter \cite{Kokkelmans2002a,Kohler2003a}, the condensate and molecular dynamics in Feshbach crossing experiments \cite{Yurovsky2003a,Kohler2004a,Goral2004a}, as well as the many-body dynamics near a photoassociative resonance \cite{Gasenzer2004a,Gasenzer2004b}.

Note that the ``diagonal'' part of the equation of motion (\ref{eq:EOMFmHFB}) for $F_m$, i.e., the term in square brackets on the left-hand side, receives an additional term $V(\xv-\yv,t)\,\widetilde{m}(\xv,\yv;t)$ when the equation is reexpressed in terms of the density matrices $\tilde n$ and $\tilde m$:
\begin{align}
\label{eq:VFmTerm}
  &\big[i\hbar\partial_{t}-H_\mathrm{1B}(\xv)-H_\mathrm{1B}(\yv)
       -V(\xv-\yv,t)\big]\tilde{m}(\xv,\yv;t)
  \nonumber\\
  &\ =\ V(\xv-\yv,t)\phi(\xv,t)\phi(\yv,t)
   + ...
\end{align} 
Hence, the diagonal part of the dynamic equation for $\tilde{m}$ contains the two-body Hamiltonian $H_\mathrm{2B}(x,y)=H_\mathrm{1B}(x)+H_\mathrm{1B}(y)+V(x-y)$ instead of the free Hamiltonian without $V$.
It was shown in Ref.~\cite{Kohler2002a} that this is crucial when deriving the Gross-Pitaevskii equation (GPE) from a many-body Hamiltonian with a non-local coupling $V(x-y)$, in the limit where the gas is weakly interacting, i.e., dilute, and the interaction strength $g$ is constant in time.
In the GPE, this coupling constant $g$ multiplies the term non-linear in $\phi$ and is defined in terms of the $s$-wave scattering length $a$ as $g=4\pi\hbar^2a/m$.
The reason for this is that $g$ is equal to the zero-energy limit  $g=(2\pi\hbar)^3\lim_{p\to0}\langle\pv|T_\mathrm{2B}(p^2/m+i0)|\pv\rangle$ of the quantum-mechanical two-body transition ($T$)-matrix
\begin{align}
\label{eq:T2B}
   \langle\pv|T_\mathrm{2B}(E)|\pv'\rangle 
   &= \langle\pv|V(1 + G_\mathrm{2B}(E) V)|\pv'\rangle.
\end{align} 
Without the term proportional $V$ in Eq.~(\ref{eq:VFmTerm}), the full two-body Greens function $G_\mathrm{2B}=(E-H_{\mathrm{2B}})^{-1}$ in the $T$-matrix (\ref{eq:T2B}) would be replaced by the free Greens function $G_{\mathrm{2B},0}=(E-H_{\mathrm{2B}}+V)^{-1}$, see Ref.~\cite{Kohler2002a}.
As a consequence, the coupling $g$ involves only the second-order Born approximation $T_\mathrm{2B}^\mathrm{2.Born}(E)=V(1 + G_\mathrm{2B,0}(E) V)$ of the $T$-matrix, which, in general, does not give the $s$-wave scattering length in the zero-energy limit.
This also shows that the non-local bare couplings $V(r)$ at different interatomic distances $r$ do not form a suitable set of small parameters for a perturbative expansion.

In summary, taking into account the double-bubble diagram as the single contribution to $\Gamma_2$, yields the closed set of Hartree-Fock-Bogoliubov many-body equations for normal-ordered equal-time one- and two-point correlation functions, i.e., the mean field and the normal and anomalous density matrices.
Moreover, starting with a non-local (bare) coupling $V(x-y)$, the full GPE requires contributions from the double-bubble diagram, i.e., beyond one-loop order.
On the other side, when starting with a local coupling $V(x-y)=g\delta(x-y)$, the underlying assumption is that the Hamiltonian defines an effective theory which is valid only at low collision energies.
In such a theory, the GPE directly results as the ``tree-level'' approximation, where all contributions from two- and higher $n$-point functions are neglected.

We finally remark that the spectral functions $\rho_i^{\ j}$ do not appear in the HFB equations for the equal-time correlators.
Hence, the HFB equations remain invariant for vanishing commutators of the Bose fields and can therefore be regarded as fully classical.

\section{Non-perturbative expansion} 
\label{sec:Nonpert}
The dynamics of non-dilute ultracold Bose gases can not, in general, be described using a loop or coupling expansion.
In three spatial dimensions, this is the case, for instance, when the mean interparticle distance is considerably smaller than the scattering length, $\eta=na^3\gg1$.
This means that the approximation needs to be based on an expansion parameter different from the interaction strength. 

An expansion in powers of the inverse $1/\cN$ of the number of field components provides a controlled expansion parameter which is not based on weak couplings.
It can be applied to describe physics characterized by large fluctuations, such as encountered near second-order phase transitions \cite{Alford2004a}, or for extreme non-equilibrium phenomena such as parametric resonance \cite{Berges2003b} or spinodal decomposition dynamics \cite{Arrizabalaga2004a}. 
For the latter cases a 2PI coupling or loop expansion is not applicable.

In the following sections we develop such a scheme for a single-component Bose-gas, on the basis of an expansion in inverse powers of the number $\cN$ of field components.
For detailed studies of such an expansion in the context of relativistic quantum field theory confer Refs.~\cite{Berges2002a,Aarts2002b}.
The method can be applied to bosonic or fermionic theories alike if a suitable field number parameter is available, and we exemplify it here for the case of a scalar $O(\cN)$-symmetric theory.
Again, we point out that the apparently rapid convergence of the 2PI expansion even for small values of ${\cal N} = 2$ is crucial for our approach~\cite{Aarts2002a,Alford2004a}.

We will first provide the relevant terms in a $1/\cN$ expansion of the 2PI
effective action to next-to-leading order and use these to derive the dynamic equations.

\subsection{$1/\cN$ expansion of the 2PI effective action} 
\label{sec:2PIN}
In the following we consider a systematic non-perturbative approximation scheme for the 2PI effective action. 
It classifies its contributions according to their scaling with powers of $1/\cN$, where $\cN$ denotes the number of field components.
We will assume that no further $U(1)$ symmetry characterizes the theory, i.e., we choose all $\cN$ field components $\Phi_i$ to be real.
For $\cN=2$ the $O(2)$-symmetric theory for real fields which we again denote as $\Phi_1$ and $\Phi_2$, is equivalent to the $U(1)$-symmetric theory considered in Section \ref{sec:PT} for the complex field $\Psi = (\Phi_1 + i \Phi_2)/\sqrt{2}$.

The 2PI $1/\cN$ approximation scheme has been derived for a relativistic field theory with local interactions in Refs.~\cite{Berges2002a,Aarts2002b}.
Before we proceed, we would like to point out that the 2PI $1/\cN$ scheme is different from the standard 1PI $1/N$ resummation which involves the free instead of the full 2-point Greens function.
Here we consider a non-relativistic field theory with the classical action given by Eq.~(\ref{eq:ClassAct}).
Recall that {\it connected} $n$-point Greens functions with $n\ge2$ are equivalent to ordinary $n$-point functions for the fluctuation field $\tilde\Phi_i(x)$ obtained by substracting the mean field from the field operator, $\Phi_i(x)=\phi_i(x)+\tilde{\Phi}_i(x)$.
Hence, the diagrammatic expansion of the 2PI effective action part $\Gamma_2$ can be derived by use of the non-local effective interaction ($V(x-y) \equiv V(y-x)$)
\begin{align}
\label{eq:Sinttilde}
  \tilde{S}_{\rm int}[\phi,\varphi] 
  =& - \frac{1}{4\cN} \int_{xy} V(x-y) \Big[ 
  2\phi_i(x)\tilde{\Phi}_j(y) \tilde{\Phi}_j(y)\tilde{\Phi}_i(x)
  \nonumber\\
   & \qquad\qquad\quad
   +\tilde{\Phi}_i(x)\tilde{\Phi}_j(y) \tilde{\Phi}_j(y)\tilde{\Phi}_i(x)\Big],
\end{align}
which is obtained from Eq.~(\ref{eq:Sint}) by shifting $\Phi_i(x) \to \phi_i(x) + \tilde{\Phi}_i(x)$ and collecting all terms cubic and quartic in the fluctuating field $\tilde{\Phi}_i(x)$. 
One observes that, in the presence of a non-vanishing macroscopic field $\phi_i(x)$, there exists an effective three-vertex in addition to the four-vertex.

The $1/\cN$ classification scheme is based on invariants under $O(\cN)$ rotations which parametrize the 2PI diagrams contributing to $\Gamma[\phi,G]$. 
The 2PI effective action is, like the underlying quantum action $S[\Phi]$, a singlet under $O(\cN)$ rotations.
As mentioned already in Section \ref{sec:NumCons}, only the $O(\cN)$ invariant $\tr\, \phi\phi \equiv \phi^2 = \phi_i \phi_i \sim \cN$ can be constructed from the mean field alone. 
All other functions of $\phi$ and $G$, which are singlets under $O(\cN)$, can be built from the irreducible invariants given in Eq.~(\ref{eq:ONinvariants}) above.
From these,  for a given $\cN$, only the invariants with $n \le \cN$ are linearly independent --- there can not be more invariants than fields
\footnote{For example, $\cN=2$ gives 5 independent invariants, in accordance with the two field components of $\phi$ and the three independent components of the Hermitian matrix $G$.}. 
In particular, for the next-to-leading order approximation one finds that only invariants with $n \le 2$ appear, which makes it very suitable for practical calculations.
 
The factors of $\cN$ in a given graph contributing to $\Gamma[\phi,G]$ have two origins: 
Each irreducible invariant is taken to scale proportional to $\cN$ since it contains exactly one trace over the field indices, while each vertex provides a factor of $1/\cN$, cf.~Eq.~(\ref{eq:Sinttilde}).
The factors of $\cN$ arising from the irreducible invariants (\ref{eq:ONinvariants}) correspond to the number of closed lines following the field indices in a diagram, plus the number of lines connecting two insertions of the classical field $\phi$.

\begin{figure}[tb]
\begin{center}
\includegraphics[width=0.45\textwidth]{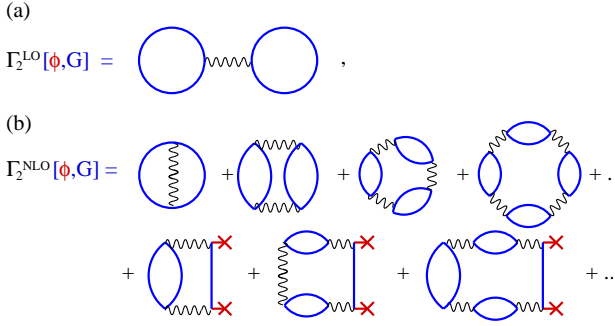}
\end{center}
\vspace*{-3ex}
\caption{
(Color online) Diagrammatic representation of the diagrams contributing, in leading order (LO) and next-to-leading order (NLO) of the $1/\cN$-expansion, to the 2PI part $\Gamma_2[\phi,G]$ of the 2PI effective action, cf.~Eqs.~(\ref{eq:Gamma2LO}) and (\ref{eq:Gamma2NLO}).
At each vertex, it is summed over double field indices and integrated over double space-time variables.
For an explanation of the symbols see the caption of Fig.~\protect\ref{fig:Gamma2PI1loop}.
}
\label{fig:Gamma2NLO1Nnonloc}
\end{figure}
The expression (\ref{eq:2PIEAexp}) for the 2PI effective action contains, besides the classical action, the one-loop contribution proportional to $\Tr(\ln G^{-1} + G_0^{-1}[\phi] G)$, and a $\Gamma_2[\phi,G]$ which contributes if higher loops are taken into account.
The one-loop term contains both leading order (LO) and next-to-leading-order (NLO) contributions in an expansion in powers of $1/\cN$.  
The logarithmic term corresponds, in absence of other terms, to the free-field effective action and scales proportional to $\cN$. 
To separate the LO and NLO contributions at the one-loop level consider the second term $\Tr(G_0^{-1}[\phi] G)$. 
From the form of the classical propagator, cf.~Eqs.~(\ref{eq:invfreeProp}), (\ref{eq:G0invexpl}), one observes that it can be decomposed into a term proportional to $\tr(G) \sim \cN$ and terms $\sim (V/\cN)[\tr(\phi\phi)\,\tr(G) + 2\,\tr(\phi\phi G)]$.
This can be seen as the sum of two ``2PI one-loop graphs'' with field insertion $\sim \phi_i\phi_i$ and $\sim \phi_i\phi_j$, respectively, as shown in Fig.~\ref{fig:Gamma2PI1loop}.
Counting the factors of $\cN$ coming from the traces and the prefactor, one finds that only the term $\sim \tr(\phi\phi)\,\tr(G)$ contributes at LO, i.e., it scales proportional to $\cN$, while the term $\sim \tr(\phi\phi G)$ is of NLO.

The LO contribution to $\Gamma_2[\phi,G]$ consists of only one two-loop graph, which is $\phi$-independent and corresponds to the first contribution to $\Gamma_2^\mathrm{HFB}$ in Eq.~(\ref{eq:Gamma2HFB}) (cf.~Fig.~\ref{fig:Gamma2HFBnonloc}):
\begin{align}
  \label{eq:Gamma2LO} 
  \Gamma_2^\mathrm{LO}[G] 
  &= - \frac{1}{4\cN} 
  \int_{xy} G_{ii}(x,x) V(x-y) G_{jj}(y,y) \, .  
\end{align} 
It is shown diagrammatically in Fig.~\ref{fig:Gamma2NLO1Nnonloc}a.
In NLO (Fig.~\ref{fig:Gamma2NLO1Nnonloc}b) there is an infinite series of contributions which can be summed up analytically \cite{Berges2002a,Aarts2002b}:
\begin{align}
\label{eq:Gamma2NLO} 
  &\Gamma_2^\mathrm{NLO}[\phi,G] 
  = \frac{i}{2} \Tr \ln (B[G]) 
  \nonumber\\
  &\qquad
  + \frac{i}{\cN}\int_{xyz} 
  I(x,z;G)V(z-y)\,\phi_i(x) \, G_{ij} (x,y) \, \phi_j (y).
\end{align} 
Here,
\begin{align}
\label{eq:BxyG} 
  B(x,y;G) 
  &=  \delta(x-y)
  \nonumber\\
  &\quad + i \frac{1}{\cN} \int_z V(x-z)\, G_{ij}(z,y)G_{ij}(z,y),
\end{align} 
with
\begin{align}
\label{eq:logexpansion}
  &\Tr \ln [B(G)]
  \nonumber\\ 
  &\quad= \int_{x} 
  \left( i\frac{1}{\cN} \int_y V(x-y)\,G_{ij}(y,x)G_{ij}(y,x) \right)
  \nonumber\\
  &\qquad- 
  \frac{1}{2} \int_{xy}
  \left( i\frac{1}{\cN} \int_z V(x-z)\,G_{ij}(z,y)G_{ij}(z,y) \right)
  \nonumber\\
  &\qquad\qquad\quad \times
  \left( i \frac{1}{\cN} \int_w V(y-w)\,G_{kl}(w,x)G_{kl}(w,x) \right)
  \nonumber\\
  &\qquad+ \ldots,
  \\  
\label{eq:Ifunc}
  &I (x,y;G) 
  = \frac{1}{\cN} \int_z V(x-z) G_{ij}(z,y) G_{ij}(z,y)
  \nonumber\\
  &\qquad - i\, \frac{1}{\cN} \int_{z w}\, I (x,w;G) V(w-z)
   G_{ij}(z,y) G_{ij}(z,y)
  \nonumber\\
  &\quad = \frac{1}{\cN}\int_z\Lambda(x,z;G)
   G_{ij}(z,y) G_{ij}(z,y), 
  \\
\label{eq:WxyG}
  &\Lambda(x,y;G)
  = V(x-y)-i\int_z I(x,z;G)\,V(z-y),  
\end{align}
%
\begin{figure}[tb]
\begin{center}
\includegraphics[width=0.45\textwidth]{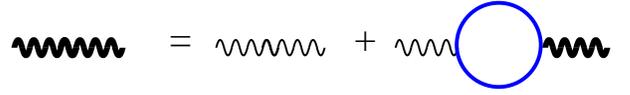}
\end{center}
\vspace*{-3ex}
\caption{
(Color online) Recursive definition of the resummed local interaction function $\Lambda(x,y;G)$, cf.~Eqs.~(\ref{eq:WxyG}) and (\ref{eq:Ifunc}), in NLO of the $1/\cN$ expansion.
The thick wiggly line represents $\Lambda$, while $G$ is, again, depicted as a thick (blue) line, and $V(x-y)$ as a thin wiggly line.
Colors are seen in the online version.
}
\label{fig:DysonLambda}
\end{figure}
The diagrammatic representation of $\Lambda$ is shown in Fig.~\ref{fig:DysonLambda}.
The first term on the rhs of Eq.~(\ref{eq:logexpansion}) corresponds to the two-loop graph with the index structure exhibiting one trace such that the contribution scales as $\tr(G^2)/\cN \sim \cN^0$.
This graph is the first in the expansion of $\Gamma_2^\mathrm{NLO}$ shown in Fig.~\ref{fig:Gamma2NLO1Nnonloc}b. 
One observes that each additional contribution scales as well proportional to $[\tr(G^2)/\cN]^n \sim \cN^0$ for all $n \ge 2$. 
Thus all terms contribute at the same order.

The functions $I(x,y;G)$ and the inverse of $B(x,y;G)$ are related by~\cite{Berges2002a,Aarts2002b}
\begin{align}
   B^{-1}(x,y;G) = \delta(x-y) - i I (x,y;G),
\label{eq:Binverse}
\end{align}
which follows from convoluting (\ref{eq:BxyG}) with $B^{-1}$ and using Eq.~(\ref{eq:Ifunc}). 
We note that $B$ and $I$ do not depend on $\phi$, and
\begin{align}
\label{eq:Gamma2toNLO}
  \Gamma_2[\phi,G]=\Gamma^\mathrm{LO}_2[G]+\Gamma^\mathrm{NLO}_2[\phi,G]+... 
\end{align} 
is only quadratic in $\phi$ at NLO.
It was shown in Ref.~\cite{Aarts2002a} that invariants containing more than two three-vertices, i.e., field insertions, are two-particle reducible and therefore can not contribute to $\Gamma_2$.
Furthermore, it was shown that graphs containing other invariants, e.g.~$\tr(G^3)$ are of higher order $\sim\cO(1/\cN)$.

\subsection{Dynamic equations} 
\label{sec:EOMNLO}
In this section we employ the stationarity conditions (\ref{eq:2PIStatCondPhi}) and (\ref{eq:2PIStatCondG}) to derive the equations of motion for the mean field $\phi$ and the two-point function $G$ to NLO in the $1/\cN$ expansion of the 2PI effective action, Eq.~(\ref{eq:Gamma2toNLO}).
In the following, we restrict ourselves again to a single complex Bose-field, $\cN=2$, which, in order to simplify the equations, we describe by its rescaled real and imaginary parts $\Phi=(\Phi_1+i\Phi_2)/\sqrt{2}$, Eq.~(\ref{eq:Phi1Phi2}).
In this representation all fields occuring in the equations will be real-valued.
We can take over all equations from earlier sections by lowering all indices and identifying $\phi=\phi_1$, $\phi^*=\phi_2$.
The equation of motion for $\phi_i(x)$ can then be read off Eqs.~(\ref{eq:EOMphi}) and (\ref{eq:Gamma2NLO}):
\begin{align}
\label{eq:EOMphiNLO1N}
   &\big[-i\sigma_{2,ij}\partial_{x_0}-H_\mathrm{1B}(\xv)\big]\,
    \phi_i(x)
   \nonumber\\
   &\ = \frac{1}{2}\int_{y}\,V(x-y)\Big[
     \Big(\phi(y)^2+G_{kk}(y,y)\Big)\phi_i(x)
   \nonumber\\
   &\qquad
    +\Big(G_{ij}(x,y)+G_{ji}(y,x)\Big)\phi_j(y)\Big]
   \nonumber\\
   &\quad
   -i\int_{y,z}I(x,z;G)V(z-y)\,G_{ij}(x,y)\phi_j(y),
\end{align} 
where we have used $G_{ij}(x,y)=G_{ji}(y,x)$ and the representation of the inverse free propagator given in Eq.~(\ref{eq:invfreeProp}).

To derive the dynamic equation (\ref{eq:EOMG}) for $G_{ij}$ in NLO of the $1/\cN$ expansion, we need 
\begin{align}
\label{eq:funcdIndG}
  \frac{\delta I(u,v)}{\delta G_{ij}(x,y)}
  &= -i\int_{wz}B^{-1}(u,w)\frac{\delta B(w,z)}{\delta G_{ij}(x,y)}
     B^{-1}(z,v)
  \nonumber\\
  &= \frac{1}{2}\int_w B^{-1}(u,w)
     \Big[V(w-x)B^{-1}(y,v)
  \nonumber\\
  &\qquad
     +V(w-y)B^{-1}(x,v)\Big]G_{ij}(x,y),
\end{align} 
where we have used Eqs.~(\ref{eq:BxyG}) and (\ref{eq:Binverse}) and suppressed the functional argument $G$ in $B$ and $I$.
We use Eqs.~(\ref{eq:Gamma2LO}), (\ref{eq:Gamma2NLO}) and (\ref{eq:funcdIndG}) to derive the self energy $\Sigma_{ij}(x,y)$, Eq.~(\ref{eq:Sigma}), in NLO. 

We then express the dynamic equations for $\phi_i$ and $G_{ij}$ in terms of
the generalized ``mass'' matrix
\begin{align}
\label{eq:Mij}
  &M_{ij}(x,y;\phi,G)
  = \delta_{ij}\delta_{\cal C}(x-y)
  \Big[H_\mathrm{1B}(x)
  \nonumber\\ 
  &\qquad
   + \frac{1}{2}\int_z V(x-z)\Big(\phi_k(z)\phi_k(z)+G_{kk}(z,z)\Big)\Big]
  \nonumber\\ 
  &\qquad
   + V(x-y)\Big(\phi_i(x)\phi_j(y)+G_{ij}(x,y)\Big)
\end{align} 
and the non-local self-energy
\footnote{We divide the self-energy into a `local' and a `non-local' part, $\Sigma=\Sigma^{(0)}+\overline{\Sigma}$. For a local coupling $V(x-y)\equiv V\delta(x-y)$, the contribution $\Sigma^{(0)}$, which we include in $M$, would be a local function, $\Sigma^{(0)}(x,y)\equiv\Sigma^{(0)}(x)\delta(x-y)$ \protect\cite{Berges2005a}.}
\begin{align}
\label{eq:Sigmabar}
  \overline{\Sigma}_{ij}(x,y;\phi,G)
  &= -i[\Lambda(x,y;G)-V(x-y)]
  \nonumber\\
  &\qquad\times
  \Big[\phi_i(x)\phi_j(y)+G_{ij}(x,y)\Big]
  \nonumber\\
  &\quad -P(x,y;\phi,G)\,G_{ij}(x,y).
\end{align} 
Here, the function $P$ is defined as
\begin{align}
\label{eq:P}
  &P(x,y;\phi,G)
   = \int_{vw} \Lambda(x,v;G)\,
  H(v,w;\phi,G)\,
  \Lambda(w,y;G),
\end{align} 
with
\begin{align}
\label{eq:H}
  H(x,y;\phi,G)
  &= \phi_i(x)G_{ij}(x,y)\phi_j(y),
\end{align} 
where, as usual, it is summed over double field indices.

\begin{figure}[b]
\begin{center}
\includegraphics[width=0.45\textwidth]{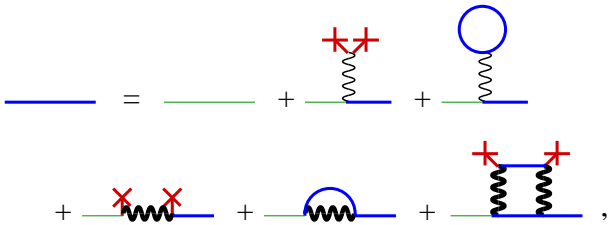}
\end{center}
\vspace*{-3ex}
\caption{
(Color online) Diagrammatic representation of the Schwinger-Dyson equation for the full propagator $G(x,y)$, in NLO of the $1/\cN$ expansion.
A thin (green) line represents the classical free propagator $D(x,y)$, cf.~Eq.~(\ref{eq:invfreeProp}).
The (red) crosses represent classical fields $\phi(x)$, the thick (blue) lines the full propagator $G(x,y)$.
The thick wiggly lines represent the resummed nonlocal interaction $\Lambda(x,y;G)$ recursively defined in Eq.~(\ref{eq:WxyG}) as illustrated in Figure \protect\ref{fig:DysonLambda}.
Each end of the solid lines as well as the external fields carry, as before, a field index and a space-time argument, which are summed/integrated over at the vertices.
}
\label{fig:SchwDysonGNLO1N}
\end{figure}

The dynamic equations then assume the compact form:
\begin{align}
\label{eq:EOMphiNLO1NwMSigma}
   &\Big[-i\sigma_{2,ij}\partial_{x_0}
         -\delta_{ij}\frac{1}{2}\int_zV(x-z)\phi_k(z)\phi_k(z)
    \Big]\phi_j(x)
   \nonumber\\
   &\quad = \int_{y}\Big[M_{ij}(x,y;\phi\equiv0           ,G)
   +i\overline{\Sigma}_{ij}(x,y;\phi\equiv 0;G)\Big]\,\phi_j(y),
   \\
\label{eq:EOMGNLO1N}
   &-i\sigma_{2,ik}\partial_{x_0}G_{kj}(x,y)
   -i\delta_{\cC}(x-y)\delta_{ij}
   \nonumber\\
   &\quad 
     = \int_z\Big[M_{ik}(x,z;\phi,G)
                  +i\overline{\Sigma}_{ik}(x,z;\phi,G)\Big]\,G_{kj}(z,y).
\end{align} 
The dynamic equation (\ref{eq:EOMGNLO1N}) for the propagator, multiplied from the left by the free classical propagator $D$, can be diagrammatically represented as the Dyson series shown in Fig.~\ref{fig:SchwDysonGNLO1N}.

Note that one recovers,  from Eqs.~(\ref{eq:EOMphiNLO1NwMSigma}) and (\ref{eq:EOMGNLO1N}), the dynamic equations in two-loop approximation of the 2PI effective action by setting $\Lambda(x,y)\equiv V(x-y)$ in Eq.~(\ref{eq:Sigmabar}), and the Hartree-Fock-Bogoliubov equations by additionally setting $P\equiv0$, i.e., by setting $\overline{\Sigma}_{ij}\equiv 0$.

To obtain numerically tractable equations of motion we finally express all functions in terms of their spectral and statistical components defined in Eq.~(\ref{eq:GintoFandrho}) and, for $\overline{\Sigma}_{ij}$, $I$, $P$, and $H$ accordingly:
\begin{align}
\label{eq:SigmarbarFandrho}
  \overline{\Sigma}_{ij}(x,y)
  &= \Sigma^F_{ij}(x,y)
    -\frac{i}{2}\Sigma^\rho_{ij}(x,y)\,
    \mathrm{sign}_{\cal C}(x_0-y_0),
    \quad\mbox{etc.}
\end{align} 
Using, furthermore,
\begin{align}
\label{eq:dGndx0inFandrho}
   \rho_{ij}(x,y)\partial_{x_0}\mathrm{sign}_{\cal C}(x_0-y_0)/2
   &= \rho_{ij}(x,y)\delta_{\cal C}(x_0-y_0)
   \nonumber\\
   &= -i\sigma_{2,ij}\delta_{\cal C}(x-y)
\end{align} 
we obtain the dynamic equations
\begin{widetext}
\begin{align}
\label{eq:EOMphiNLO1NinFrho}
   &\Big[-i\sigma_{2,ij}\partial_{x_0}
    -\delta_{ij}\frac{1}{2}\int_zV(x-z)\phi_k(z)\phi_k(z)
    \Big]\phi_j(x)
    -\int_yM_{ij}(x,y;\phi\equiv0,F)\,\phi_j(y)
   = \int_0^{x_0}dy\,\Sigma^\rho_{ij}(x,y;\phi\equiv 0;G)\,\phi_j(y),
   \\
\label{eq:EOMFNLO1NinFrho}
   &-i\sigma_{2,ik}\partial_{x_0}F_{kj}(x,y)
    -\int_zM_{ik}(x,z;\phi,F)\,F_{kj}(z,y)
    = \int_0^{x_0}dz\,\Sigma^\rho_{ik}(x,z;\phi,G)F_{kj}(z,y)
     -\int_0^{y_0}dz\,\Sigma^F_{ik}(x,z;\phi,G)\rho_{kj}(z,y),
   \\
\label{eq:EOMrhoNLO1NinFrho}
   &-i\sigma_{2,ik}\partial_{x_0}\rho_{kj}(x,y)
    -\int_zM_{ik}(x,z;\phi,F)\,\rho_{kj}(z,y)
    = \int_{y_0}^{x_0}dz\,\Sigma^\rho_{ik}(x,z;\phi,G)\rho_{kj}(z,y),
\end{align} 
\end{widetext}
where $\int_t^{t'}dx\equiv\int_t^{t'}dx_0\int d^3x$.
The real functions $M_{ij}(x,y;\phi,F)$, $\Sigma^F_{ij}(x,y;\phi,G)$, and $\Sigma^\rho_{ij}(x,y;\phi,G)$ are all regular in $x_0$ and are explicitly provided, in terms of the fields $\phi_i(x)$, $F_{ij}(x,y)$, and $\rho_{ij}(x,y)$, in Appendix \ref{app:FuncsMandSigma}.

Note that, as discussed in Section \ref{sec:NumCons}, the effective action, to any order of the $1/\cN$ expansion, is manifestly invariant under $O(\cN)$ rotations, i.e., for $\cN=2$, the dynamic equations locally conserve, to any order in the expansion, the particle number density. 

\section{Non-equilibrium dynamics in one spatial dimension}
\label{sec:1DDynamics}
In the following we will apply the dynamic equations derived from the 2PI effective action, to next-to-leading order (NLO) in the $1/{\cal N}$ expansion, to the case of an ultracold atomic Bose gas in one spatial dimension.
A full dynamical theory of the formation of a Bose-Einstein condensate out of a gas with non-equilibrium distribution of single-particle energies, as occurring in everyday experiments, is still an open and intriguing problem in the field of atomic matter-wave physics \cite{Kohl2002a}.
As a first step towards a more complete description of the dynamics of crossing the Bose-Einstein phase transition we study the time evolution of a uniform one-dimensional Bose gas which is, initially, characterized by a non-equilibrium distribution of particle momenta. 
In this chapter, we present results for the time evolution of such a system as obtained from numerically solving the dynamical equations introduced above.

In one spatial dimension, there is no phase transition to a Bose-Einstein condensate, i.e., to a state where only the lowest energy mode is macroscopically populated. 
Consequently, there is no condensate mean field.
Nevertheless, at very low temperatures, equilibrium states with a macroscopic occupation of the lowest energy levels are possible, a phenomenon which manifests itself in approximate off-diagonal long-range order (ODLRO), i.e., in a first-order phase coherence which extends over the entire gas sample \cite{Pitaevskii2003a} but can, under specific experimental conditions, be broken more easily than in three dimensions. 
On general grounds one expects that, starting with a non-equilibrium momentum distribution, the system is driven towards a state described by a Bose-Einstein equilibrium distribution with a suitable dispersion relation.  

For vanishing mean field, the NLO dynamical equations for the statistical and spectral correlation functions $F$ and $\rho$, respectively, are given in Eqs.~(\ref{eq:EOMFNLO1NinFrho}) and (\ref{eq:EOMrhoNLO1NinFrho}), with $\phi\equiv0$, and space-time variables $x=(x_0,x_1)$, etc. 
We have chosen a local coupling 
\begin{align}
\label{eq:1Dcoupling}
  V(x-y) = g_\mathrm{1D}\delta(x-y),
\end{align} 
where the coupling constant in one dimension, $g_\mathrm{1D}$, is related to the dimensionless $\gamma$ parameter and the total line-density of atoms $n_1$ by \cite{Pitaevskii2003a} 
\begin{align}
\label{eq:1Dcouplitogamma}
  g_\mathrm{1D}=\frac{\hbar^2n_1}{m}\gamma.
\end{align} 
For the approximation in Eq.~(\ref{eq:1Dcoupling}) to be justified we assume that bound states of two and more atoms are unlikely to be formed, that three-body collisions can be neglected, and that the interaction potential $V$ is constant in time.

We solve the corresponding system of equations in momentum space,
%
\begin{align}
\label{eq:EOMFpNLO1NinFrho}
   &\Big(-i\sigma_{2,ik}\partial_{t}
    -M_{ik}(t,p;F)\Big)\,F_{kj}(t,t';p)
   \nonumber\\
   &\qquad\qquad
    = \int_0^{t}dt''\,\Sigma^\rho_{ik}(t,t'';p;G)F_{kj}(t'',t';p)
   \nonumber\\
   &\qquad\qquad\quad
     -\int_0^{t'}dt''\,\Sigma^F_{ik}(t,t'';p,G)\rho_{kj}(t'',t';p),
   \\
\label{eq:EOMrhopNLO1NinFrho}
   &\Big(-i\sigma_{2,ik}\partial_{t}
    -M_{ik}(t,p;F)\Big)\,\rho_{kj}(t,t';p)
   \nonumber\\
   &\qquad\qquad
    = \int_{t'}^{t}dt''\,\Sigma^\rho_{ik}(t,t'';p;G)
                          \rho_{kj}(t'',t';p),
\end{align} 
%
where
\begin{align}
\label{eq:Mijp}
  &M_{ij}(t,p;F)
  = \delta_{ij}
  \Big[\frac{p^2}{2m}
   + \frac{g_\mathrm{1D}}{2}\int_k F_{ii}(t,t;k)\Big]
  \nonumber\\ 
  &\qquad
   + g_\mathrm{1D}\int_k F_{ij}(t,t;k).
\end{align} 
The non-local self-energies $\Sigma^{F,\rho}_{ij}(t,t';p)$, to NLO in the $1/{\cal N}$ expansion, are given in Appendix \ref{app:FuncsMandSigma}.

With initial values for $F_{ij}(0,0;p)$, $\rho_{ij}(0,0;p)$, the above coupled system of integro-differential equations with first-order time derivative yields the time evolution of the correlation functions, in particular, for $t=\tau$, of the momentum distribution  
\begin{align}
\label{eq:npoft}
  n(t,p) = \frac{1}{2}\Big(F_{11}(t,t;p)+F_{22}(t,t;p)-1\Big).
\end{align} 
As initial conditions we chose a Gaussian distribution
\begin{align}
\label{eq:npini}
  n(0,p) = \frac{n_1}{\sqrt{\pi}\sigma}e^{-p^2/\sigma^2}.
\end{align} 
Furthermore we chose the initial pair correlation function (\ref{eq:Fm}) to vanish,
\begin{align}
\label{eq:FmIniCond}
   0&=\langle\hat{\Psi}(p,t)\hat{\Psi}(p,t)\rangle_c
   \nonumber\\
   &=\ \frac{1}{2}\Big(F_{11}(t,t;p)-F_{22}(t,t;p)\Big)
    +iF_{12}(t,t;p),
\end{align} 
for $t=0$, in accordance with total atom number conservation at non-relativistic energies
\footnote{
At non-relativistic energies, the total number of atoms in a certain finite system $A$ is fixed, say it is $N$. Then, for any subsystem $B$ of $A$, the density matrix, due to number conservation, takes the form $\rho=\sum_{N_B,N_B'}|N-N_B\rangle\langle N-N_B'|\otimes|N_B\rangle\langle N_B'|$. Hence, like the mean field $\langle\hat{\Psi}\rangle$, the expectation value $\langle\hat{\Psi}\hat{\Psi}\rangle=\Tr_B(\hat{\Psi}\hat{\Psi}\,\Tr_{A\backslash B}[\rho])$ necessarily vanishes, where $\Tr_{A\backslash B}$ denotes the trace over all basis states of the Hilbert space corresponding to the ``environment'' of $B$ in $A$.}.
\begin{figure}[tb]
\begin{center}
\includegraphics[width=0.45\textwidth]{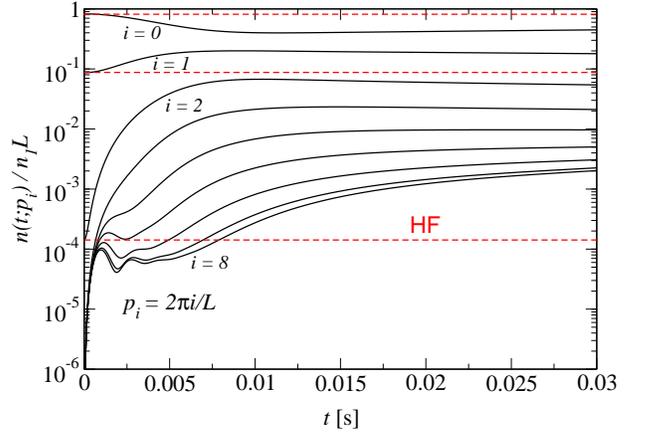}
\end{center}
\vspace*{-3ex}
\caption{
(Color online) Time evolution of occupation numbers of different momentum modes of a uniform one-dimensional gas of interacting $^{23}$Na atoms, for a dimensionless interaction parameter $\gamma=7.5\cdot10^{-4}$ and an initially Gaussian distribution, Eq.~(\ref{eq:npini}), with width $\sigma=1.3\cdot10^5\,$m$^{-1}$, according to the dynamic equations (\ref{eq:EOMFpNLO1NinFrho}), (\ref{eq:EOMrhopNLO1NinFrho}) in next-to-leading order in the $1/{\cal N}$ approximation. 
Shown are the relative populations $n(t,p_i)/n_1L$ on a logarithmic scale as a function of evolution time $t$, where $n(t,p)=F_{11}(t,t;p)-1/2$ (cf.~Eqs.~(\ref{eq:npoft}), (\ref{eq:IniValF}) and the discussion in the text), and $n_1=10^7\,$m$^{-1}$ is the total line density.
The momenta $p_i=2\pi i/L$ in the periodic box of length $L=32\,\mu$m are labelled by their mode numbers $i$.
The lattice spacing is $a_s=2\,\mu$m such that the figure shows the entire momentum spectrum.
The dashed lines labelled by ``HF'' show the time evolution of the lowest momentum modes $i=0,1,2$ resulting from the Hartree-Fock approximation to the dynamic equations (\ref{eq:EOMFpNLO1NinFrho}), (\ref{eq:EOMrhopNLO1NinFrho}). 
In this approximation, the momentum distribution is constant in time!
The full NLO dynamics given by the solid lines shows that a quasistationary momentum distribution is reached quickly.
}
\label{fig:n1poft}
\end{figure}
As a consequence,
\begin{align}
\label{eq:IniValF}
&F_{11}(0,0;p)=F_{22}(0,0;p)=n(0,p)+1/2,
\\
&F_{12}(0,0;p)=F_{21}(0,0;p)\equiv0.
\end{align} 
Indeed, the condition (\ref{eq:FmIniCond}) proved to be conserved in time, as required by number conservation, such that $n(t,p)=F_{11}(t,t;p)-1/2$.
The Bose commutation relations, furthermore, require that
\begin{align}
\label{eq:IniValrho}
\rho_{11}(t,t;p)=\rho_{22}(t,t;p)\equiv0&,
\\
-\rho_{12}(t,t;p)=\rho_{21}(t,t;p)\equiv1&.
\end{align} 

As a specific example we consider an ultracold gas of $^{23}$Na atoms with a line density of $n_1=10^7\,$m$^{-1}$. 
The dimensionless interaction parameter is chosen to be $\gamma=7.5\cdot10^{-4}$ (Cf.~Eq.~(\ref{eq:1Dcouplitogamma})). 
In an experiment such an interaction parameter could be realized, for a given line density and $s$-wave scattering length, by confining the gas in the transverse direction by a sufficiently steep potential. 
Assuming a transverse harmonic potential, the chosen value for $\gamma$ corresponds to an oscillator length $\ell_\perp=\sqrt{2a/n_1\gamma}=0.88\,\mu$m, where $a=54.6\,a_\mathrm{Bohr}=2.89\,$nm is the $s$-wave scattering length of the $^{23}$Na atoms in the commonly trapped state $|F=1,m_F=-1\rangle$ \cite{Mies2000a}. 
The width of the initial momentum distribution is taken to be $\sigma=1.3\cdot10^5\,$m$^{-1}$. 
\begin{figure}[tb]
\begin{center}
\includegraphics[width=0.45\textwidth]{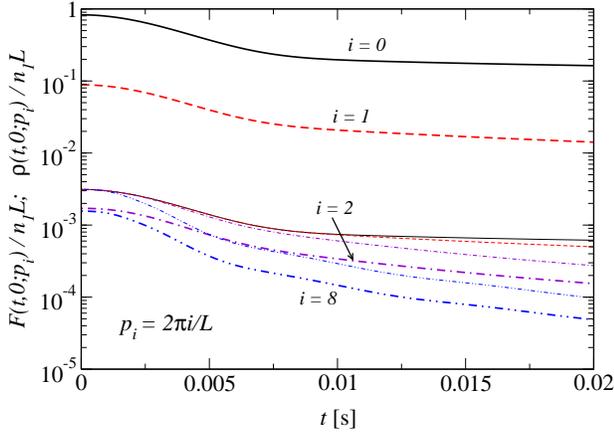}
\end{center}
\vspace*{-3ex}
\caption{
(Color online) Envelopes of the two-point statistical and spectral functions $F_{ij}(t,0;p_i)$ and $\rho_{ab}(t,o;p_i)$, $a,b\in\{1,2\}$, respectively, normalized to the total number of particles in the box $n_1L$, as a function of time $t$ for different momentum modes $p_i=2\pi i/L$. 
These envelopes have been calculated as $F(t,0;p_i)\equiv[\sum_{a=1}^2 F_{1a}(t,0;p_i)^2]^{1/2}$ and $\rho(t,o;p_i)\equiv[\sum_{a=1}^2 \rho_{1a}(t,0;p_i)^2]^{1/2}$, respectively. 
All parameters are chosen as in Fig.~\ref{fig:n1poft}.
The statistical functions $F$ are depicted by thick lines, with pattern (and color) labelling a particular momentum mode. 
The spectral functions $\rho$ are given by the correspondingly drawn thin lines.
One observes that at time $t=0$ the normalized spectral functions all go to the fixed value $1/(2n_1L)$, in accordance with the equal time commutator.
At times $t\gtrsim0.01\,$s, the respective statistical and spectral functions for a particular momentum mode show an exponential behaviour with the same decay constant $\gamma_\mathrm{damp}(p_i)$.
}
\label{fig:Frhot0Envoft}
\end{figure}

Figure \ref{fig:n1poft} shows the time evolution of the occupation numbers $n(t,p_i)$, normalized to the total number of particles in the box, $n_1L$, according to the dynamic equations (\ref{eq:EOMFpNLO1NinFrho}), (\ref{eq:EOMrhopNLO1NinFrho}) in next-to-leading order in the $1/{\cal N}$ approximation. 
The relative populations $n(t,p_i)/n_1L$ are shown on a logarithmic scale.
The momenta $p_i=2\pi i/L$ in the periodic box of length $L=32\,\mu$m are labelled by their mode numbers $i$.
The lattice spacing is $a_s=2\,\mu$m, so the figure shows the entire momentum spectrum. 

\begin{figure}[tb]
\begin{center}
\includegraphics[width=0.45\textwidth]{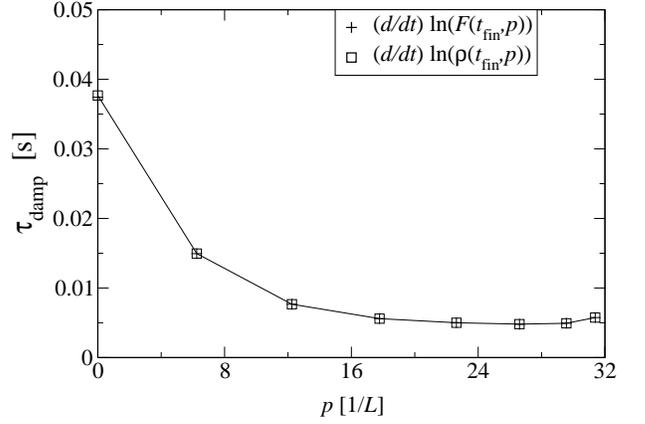}
\end{center}
\vspace*{-3ex}
\caption{
Decay time $\tau_\mathrm{damp}=\gamma_\mathrm{damp}$ derived from the slope $-\gamma_\mathrm{damp}/2$ of the envelopes of the two-point statistical and spectral functions on the logarithmic scale shown in Fig.~\ref{fig:Frhot0Envoft}, for the different momentum modes, at time $t=0.02\,$s.
$\tau_\mathrm{damp}$ is shown as a function of momentum. $p=2\sin(a_sp_i/2)/a_s$ is the momentum corresponding to the discretized Laplacian for a lattice constant $a_s$.
Within the resolution of the figure, the results for the statistical and spectral functions are identical.
$\tau_\mathrm{damp}$ is the characteristic decay time over which specific information about the initial state is lost.
This corresponds to the time when, for a particular momentum mode, the strong initial oscillations seen in Fig.~\ref{fig:n1poft} are damped out, and the systems enters the quasistationary period.
The time scale $\tau_\mathrm{damp}$ is, in the weak-coupling regime considered here, orders of magnitude smaller than the time scale $\tau_\mathrm{eq}$ for the thermalization of the system. 
}
\label{fig:slopeEnvFrhot0ofp}
\end{figure}
We point out that the dynamic equations (\ref{eq:EOMFpNLO1NinFrho}), (\ref{eq:EOMrhopNLO1NinFrho}) in the Hartree-Fock (HF) approximation, where the non-local self-energies $\Sigma^{F,\rho}_{ik}$ vanish identically, do not give any change of the initial momentum distribution in time. 
This is due to the absence of direct scattering in the HF-approximation.
The HF- as well as the LO $1/{\cal N}$ approximation suffer from an infinite number of spurious conserved quantities, which are not present in the fully interacting theory. 
These spurious constants of motion are associated to an infinite life-time of quasi-particle momentum modes, which prevent relaxation to a thermal  distribution~\cite{Bettencourt1998a,Berges2002a}.
In Figure \ref{fig:n1poft} this is indicated by the dashed lines labelled ``HF'' which give the occupation of the lowest three momentum modes $i=0,1,2$.

In contrast to this, the full NLO dynamics given by the solid lines shows that a quasistationary momentum distribution is reached after a short evolution time
\footnote{Other than in a relativistic theory, where $1\to3$ on-shell reactions with pair production lead to thermalization \protect\cite{Berges2003a}, at non-relativistic energies strict number conservation only admits $2\to2$-vertices. Moreover, due to the restricted phase space in 1+1 dimensions, momentum conservation leaves all momenta invariant. Hence, the momentum occupations are not damped in $2\to2$ on-shell reactions, and thermalization requires a finite width of the propagator.}.
The characteristic time $\tau_\mathrm{damp}$ of the damping of initial oscillations corresponds to the time over which the detailed information about the initial state is lost.
We have checked explicitly that for different initial momentum distributions, which are peaked around a finite value of $|p|$, but have a smaller width, such that their total particle number and energy content are the same, the momentum distribution settles to the same quasistationary distribution within $\tau_\mathrm{damp}$. 

This loss of information shows up in the decrease of the two-point functions at large time differences $t-t'$, where the time dependence of its envelope becomes an exponential of $t-t'$.   
Figure \ref{fig:Frhot0Envoft} shows, for different momentum modes $p_i=2\pi i/L$, the envelopes of the two-point statistical and spectral functions $F_{ij}(t,0;p_i)$ and $\rho_{ab}(t,0;p_i)$, $a,b\in\{1,2\}$, respectively, normalized to the total number $n_1L$, as a function of time $t$. 
One observes that at times greater than about $0.01\,$s, the respective statistical and spectral functions for a particular momentum mode show an exponential behaviour 
\begin{align}
\label{eq:Frhot0damping}
   F(t,t';p) \sim \exp\{-\gamma_\mathrm{damp}(p)(t-t')/2\},
\end{align} 
with the same decay constant $\gamma_\mathrm{damp}(p_i)$.
In Figure \ref{fig:slopeEnvFrhot0ofp}, we show the corresponding time scale $\tau_\mathrm{damp}(p)=\gamma_\mathrm{damp}^{-1}(p)$ for the different momentum modes $p_i$, at time $t=0.02\,$s.  
Within the resolution of the figure, no difference between the results for the statistical and spectral functions appear. 
Comparing with Fig.~\ref{fig:n1poft} it becomes clear that $\tau_\mathrm{damp}(p)$ corresponds to the time when, for a particular momentum mode, the strong initial oscillations of the occupation number are damped out, and the systems enters the quasistationary period. 
However, the time scale $\tau_\mathrm{damp}$ is, in the weak-coupling regime considered here, orders of magnitude smaller than the time scale $\tau_\mathrm{eq}$ for the thermalization of the system, cf.~also Refs. \cite{Aarts2000a,Berges2002a,Berges2003a}.  

It is important to realize that for times smaller than $\tau_\mathrm{damp}(p)$ the statistical and spectral functions are not characterized by the same damping. 
This reflects the fact that the far-from-equilibrium dynamics is not characterized by a fluctuation-dissipation relation, which would relate the damping of $F$ and $\rho$ (see Eq.~(\ref{eq:FlucDiss})).
On the other hand, our results indicate the approximate validity of a fluctuation-dissipation relation for times larger than $\tau_\mathrm{damp}$.

\begin{figure}[tb]
\begin{center}
\includegraphics[width=0.45\textwidth]{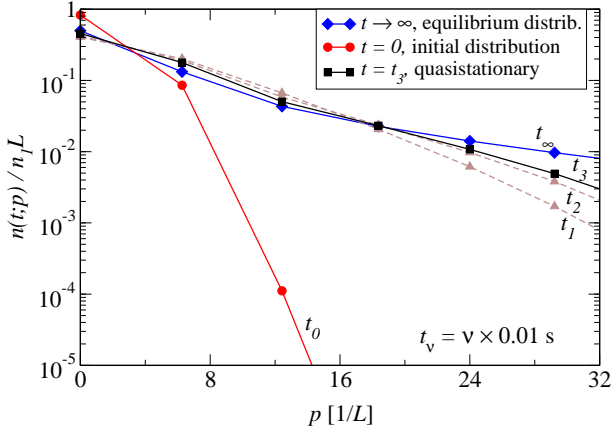}
\end{center}
\vspace*{-3ex}
\caption{
(Color online) Occupation numbers of the modes vs momentum, for consecutive time steps of the evolution of the one-dimensional gas.
Shown are the normalized populations $n(t,p_i)/n_1L$ on a logarithmic scale as a function of momentum $p$, for a box length $L=32\,\mu$m and a lattice spacing $a_s=1.73\,\mu$m.
The (red -- color online) circles show the far-from-equilibrium initial Gaussian momentum distribution, Eq.~(\ref{eq:npini}), the (black) squares the momentum distribution at $t=t_3=0.03\,$s, and the (brown) triangles the distribution at times $t_\nu=\nu\times0.01\,$s, $\nu=1,2$ (their linear interpolation being added as a guide to the eye).
The latter, quasistationary distribution developed after a short time is close to the thermal equilibrium distribution, which is expected, however, to be reached only much later. 
The (blue) diamonds represent an estimate for the true equilibrium distribution corresponding to the initial density and energy.
}
\label{fig:n1pofp}
\end{figure}

The occupation numbers of the different momentum modes are shown in Figure \ref{fig:n1pofp}, for the same parameters as in Fig.~\ref{fig:n1poft}, laid out vs. the momentum $p=2\sin(a_sp_i/2)/a_s$  corresponding to the discretized Laplacian for a lattice constant $a_s$. 
The relative populations $n(t,p_i)/n_1L$ are again shown on a logarithmic scale such that a Gaussian corresponds to an inverted parabola. 

The (red) circles show the far-from-equilibrium initial Gaussian momentum distribution, Eq.~(\ref{eq:npini}), the (black) squares the momentum distribution at $t=t_3=0.03\,$s (their linear interpolation being added as a guide to the eye).
The (brown) triangles interpolated by dashed lines which correspond to intermediate times $t_1=0.01\,$s and $t_2=0.02\,$s demonstrate that the distribution quickly converges. 
No significant change of the momentum distributions have been found for a larger momentum cutoff $1/a_s$.
From the initial momentum distribution we have calculated the initial total energy   
\begin{align}
\label{eq:Eini}
  E_0
  &= \frac{1}{2}\sum_{p_i}\Big\{\Big(\frac{p^2}{2m}
     +\frac{g_\mathrm{1D}}{4}\sum_{k_i}F_{aa}(0,0;k_i)\Big)\delta_{ab}
   \nonumber\\
  &\qquad
     +\frac{g_\mathrm{1D}}{2}\sum_{k_i}F_{ab}(0,0;k_i)\Big\}F_{ab}(0,0;p_i)
  \nonumber\\
  &= \sum_{p_i}\frac{p^2}{2m}n(0,p)+g_\mathrm{1D}n_1(n_1L+1).
\end{align} 
Given this energy and total density $n_1$, as well as the fact that the gas is weakly interacting ($n_1a\simeq0.03\ll1$) we derived an estimate for the equilibrium momentum distribution in the HFB approximation, $n_\mathrm{eq}(p)=(u_p^2+v_p^2)\{\exp[\beta(\omega(p)-\mu)]-1\}^{-1}+v_p^2$, where $\mu$ is the chemical potential, $\beta$ the inverse temperature, $u_p=\{1-[(p^2/2-\omega(p))/\gamma n_1^2+1]^2\}^{-1/2}$ and $v_p=\{[(p^2/2-\omega(p))/\gamma n_1^2+1]^{-2}-1\}^{-1/2}$ the HFB coefficients, and $\omega(p)$ is defined by the Bogoliubov dispersion relation 
$\omega(p)=(p/m)\sqrt{\gamma n_1^2+p^2/4}$.
This estimate was obtained using number and energy conservation, i.e., $n_1L=\sum_{p_i}n_\mathrm{eq}(p_i)$ and $E_0=\sum_{p_i}\omega(p)n_\mathrm{eq}(p_i)$, respectively.  
The resulting equilibrium distribution is shown in Fig. \ref{fig:n1pofp} by the (blue) diamonds.
Hence, as can be seen in the Figure, the quasistationary momentum distribution reached at the end of the evolution shown in Fig.~\ref{fig:n1poft} is already close to the expected equilibrium distribution. 

\begin{figure}[tb]
\begin{center}
\includegraphics[width=0.45\textwidth]{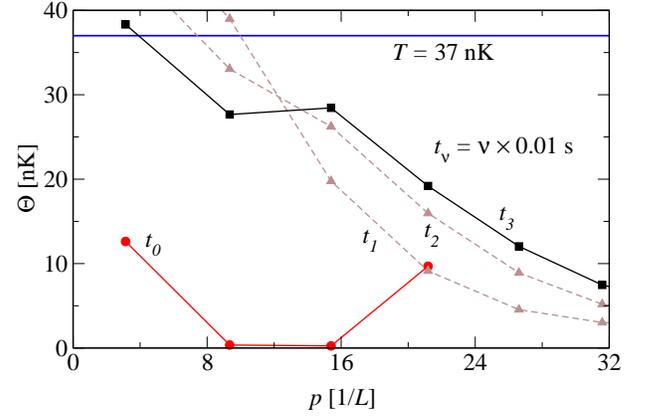}
\end{center}
\vspace*{-3ex}
\caption{
(Color online) The temperature variable $\Theta(t_n,p)$ derived, using Eq.~(\ref{eq:Tfromnt}), from the non-equilibrium momentum distribution at times $t_\nu=\nu\times0.01\,$s, $\nu=0,1,2,3$ shown in Fig.~\ref{fig:n1pofp}.
Since a quadratic, i.e., free dispersion relation, $\omega(p)=p^2/2m$, has been assumed, the data indicates the curvature of inverted parabolae locally fitted to the momentum distributions at different times $t_n$ and momenta $p_i$.
For $t\to\infty$, the values of $\Theta$ are expected to approach the temperature $T=37\,nK$ of the equilibrium distribution indicated by the (blue) solid line.
}
\label{fig:Ttofp}
\end{figure}
In experiment, a usual procedure for determining the temperature of an ultracold gas is to fit a Gaussian to the observed momentum distribution of the atoms. 
It is therefore interesting to compare the temperature one would obtain for the non-equilibrium distribution during the quasistationary period with the temperature of the anticipated equilibrium distribution indicated by the diamonds in Fig.~\ref{fig:n1pofp}. 
For sufficiently large momenta $p$ the dispersion relation of the weakly interacting atoms will be close to that of free particles: 
$\omega(p)\simeq p^2/2m$.
Hence, assuming a Bose-Einstein distribution we define an effective temperature variable as
\begin{align}
\label{eq:Tfromnt}
  &\Theta(t,p) = \Big\{\frac{mk_B}{p}\frac{\ptd}{\ptd p}\ln\left[1+\frac{1}{n(t,p)}\right]\Big\}^{-1}.
\end{align} 
This is shown, as a function of $p$, for different evolution times $t_\nu=\nu\times0.01\,$s, $\nu=0,...,3$, in Figure \ref{fig:Ttofp}, and compared to the temperature $T$ of the anticipated equilibrium distribution (diamonds in Fig.~\ref{fig:n1pofp}).
For large $p$, where the dispersion is approximately quadratic, the value for $\Theta$ derived from the equilibrium distribution is far from the equilibrium temperature $T=37\,$nK.
However, at small momenta, where the occupation numbers are large and can more easily measured in experiment, the order of magnitude of the derived temperature variable $\Theta(p)$ coincides with that of the anticipated temperature. 
In contrast to this, the values $\Theta$ derived from the non-equilibrium distribution deviate significantly from the equilibrium temperature over the entire range of momenta and times. 
Hence, our results show, that during the non-equilibrium quasistationary period which sets in very quickly, the amount of information about the final temperature which can be deduced from the momentum distribution of the gas is limited.

\section{Conclusions}
\label{sec:Concl}
We have presented a dynamical many-body theory of an ultracold Bose gas which systematically extends beyond mean-field and perturbative quantum-field theoretical procedures.
The approach is valid for large interaction strengths and/or long evolution times, two conditions which have gained a great importance in experiments with ultracold atomic gases.
Examples include the Feshbach-enhanced collisions, dilute gases in one- and two-dimensional trapping geometries, the Superfluid-to-Mott-insulator transition in optical lattices, and the transition from a Bose-Einstein- to a Bardeen-Cooper-Schrieffer-type superfluid in a degenerate Fermi gas.
The methods described in this article are based on an expansion of the two-particle (2PI) effective action in powers of the inverse $1/\cN$ of the number of field components.

To our knowledge, this work represents the first full treatment of the nonrelativistic dynamics of ultracold atomic gases to next-to-leading order in the 2PI $1/\cN$-expansion.
We have reviewed the general framework for obtaining, from the 2PI effective action, (exact) dynamic equations for the lowest order many-body correlation functions, the mean field and the two-point functions which define the non-condensate and anomalous density matrices.
To be able to solve these equations we discussed different approximation schemes, which, in the present approach, are implemented at the level of the effective action.
This ensures the dynamic equations to preserve crucial symmetries such as leading to energy and number conservation.
We first described how the well-known time-dependent Hartree-Fock-Bogoliubov (HFB) equations which, in the recent past, have been used extensively in the context of non-equilibrium dynamics of ultacold atomic gases, result as the lowest-order approximation.
In the subsequent sections we have described the setting up of the non-perturbative 2PI dynamic equations to next-to-leading order in the $1/\cN$ expansion and their application to the dynamics of an ultracold quantum-degenerate gas of sodium atoms in one spatial dimension.
As a specific example we have studied the time evolution of a uniform gas with an initially far-from-equilibrium momentum distribution.
We have shown that the many-body system quickly loses the precise information about the details of its initial state and enters a quasi-stationary evolution period.
During this period, the momentum distribution is still far from equilibrium, and the drift towards the anticipated Bose-Einstein equilibrium distribution is found to be extremely slow.
Fitting a Bose-Einstein distribution during the quasi-stationary period yields effective temperatures far away from the expected final temperature.

Summarizing, we point out, that, in order to describe non-equilibrium dynamics of strongly interacting and/or dense atomic gases, an approach beyond the conventional Hartree-Fock(-Bogoliubov) approximation is required.
In fact, in the case where the system is described solely by the two-point functions, i.e., whenever the mean field vanishes, the HF dynamic equations do not describe any dynamical change in the mode occupation numbers such that thermal equilibration is not accounted for.
This is due to the fact, that the HF approximation neglects multiple
scattering and therefore conserves an infinite number of spurious conserved quantities.

We emphasize that the 2PI $1/\cN$ expansion to next-to-leading order represents an approximation scheme which is capable of describing the dynamics of strongly interacting many-body systems far from thermal equilibrium.
The study of its implications in the context of ultracold atomic gases is an exciting and demanding task for near-future research.

\acknowledgments
\noindent
We are very grateful to Szabolcs Bors\'anyi, Joachim Brand, Elmar Haller, Peter Kr\"uger, Markus Oberthaler, J\"org Schmiedmayer, and Julien Serreau for valuable discussions and suggestions, and to Werner Wetzel for support concerning computers.
This work has been supported by the Deutsche Forschungsgemeinschaft (T.G.) and the Ministerio de Educaci\'{o}n y Ciencia of Spain (M.S.), under grant EX2003-0696.

\begin{appendix}
\section{Derivation of $\Gamma[\phi,G]$ to one-loop order}
\label{app:Gamma2PI1loop}
\noindent
In Eq.~(\ref{eq:2PIEAexp}) the 2PI effective action is explicitly given up to one-loop order while all higher order contributions are denoted by $\Gamma_2[\phi,G]$.
The one-loop expression can be derived as follows:
We first note that the 2PI effective action can be viewed as the 1PI effective action 
\begin{align}
\label{eq:Gamma1PILegTrans}
  \Gamma^K[\phi]
  &= W[J,K]-\int_x\phi_i(x)J_i(x)
\end{align} 
for a theory governed by the modified classical action 
\begin{align}
\label{eq:SK}
  S^K[\Phi]
  &= S[\Phi]+\frac{1}{2}\int_{xy}\Phi_i(x)K_{ij}(x,y)\Phi_j(y),
\end{align} 
which contains an additional ``potential'' term quadratic in the fields.
For $K\equiv0$, $\Gamma^K[\phi]$ is equivalent to the conventional 1PI effective action.
The 1PI effective action $\Gamma^K[\phi]$ then assumes, to one-loop order, the well-known form 
\begin{align}
\label{eq:Gamma1PI1loop}
  \Gamma^{K(\mathrm{1loop})}[\phi]
  &= S^K[\phi] +\frac{i}{2}\mathrm{Tr}\ln\big(G_0^{-1}[\phi]-iK\big),
\end{align} 
where $iG_{0,ij}^{-1}(x,y;\phi)=\delta^2S[\phi]/\delta\phi_i(x)\delta\phi_j(y)$ is the classical inverse propagator.

Using the definition (\ref{eq:Gamma1PILegTrans}) one finds that
\begin{align}
\label{eq:dGammaKdReqdWndR}
  \frac{\delta\Gamma^K[\phi]}{\delta K_{ij}(x,y)}
  &= \frac{\delta W[J,K]}{\delta K_{ij}(x,y)}
\end{align} 
and, together with Eq.~(\ref{eq:2PILegendreCondK}), the Legendre transform of $\Gamma^K[\phi]$ with respect to $K$, i.e., the 2PI effective action can be written as
\begin{align}
\label{eq:2PIEAasLegTransofGammaK}
   \Gamma[\phi,G]
   &= \Gamma^K[\phi]
     -\int_{xy}\frac{\delta\Gamma^K[\phi]}{\delta K_{ij}(x,y)}R_{ji}(y,x)
   \nonumber\\
   &= \Gamma^K[\phi]-\frac{1}{2}\int_{xy}\phi_i(x)K_{ij}(x,y)\phi_j(y)
   \nonumber\\
   &\qquad\qquad -\frac{1}{2}\mathrm{Tr}KG.
\end{align} 
Now we insert Eq.~(\ref{eq:Gamma1PI1loop}) into Eq.~(\ref{eq:2PIEAasLegTransofGammaK}) to obtain the 2PI effective action in one-loop approximation: 
\begin{align}
\label{eq:2PIEA1loop}
   \Gamma^{\mathrm{1loop}}[\phi,G]
   &= S[\phi]
      +\frac{1}{2}\mathrm{Tr}\Big(i\ln\Big[G_0^{-1}[\phi]-iK\Big]-KG\Big).
\end{align} 
Using the one-loop relation $G^{-1}=G_0^{-1}-iK$ between the exact inverse propagator $G^{-1}$ and the classical inverse propagator of the modified action $S^K$,
\begin{align}
\label{eq:dGammaKndphiphi}
  iG_{ij}(x,y)
  &=
  \frac{\delta^2\Gamma^K[\phi]}{\delta\phi_i(x)\delta\phi_j(y)}
  \nonumber\\
  &= i\Big[G_{0,ij}^{-1}(x,y)-iK_{ij}(x,y)-\Sigma^K_{ij}(x,y)\Big],
\end{align} 
with $\Sigma^K_{ij}(x,y)$ being the proper self energy to which only one-particle irreducible diagrams contribute,
one obtains, from Eq.~(\ref{eq:2PIEA1loop}), the one-loop terms of $\Gamma[\phi,G]$ in Eq.~(\ref{eq:2PIEAexp}).

\section{The functions $M_{ij}$, $\Sigma^{(0)}$, $\Sigma^F_{ij}$, and $\Sigma^\rho_{ij}$ in NLO of the $1/\cN$ expansion}
\label{app:FuncsMandSigma}
\noindent
In Section \ref{sec:EOMNLO} we have derived the dynamic equations for the real fields $\phi_i(x)$, $F_{ij}(x,y)$, and $\rho_{ij}(x,y)$ in NLO of the $1/\cN$ expansion of the 2PI effective action, cf.~Eqs.~(\ref{eq:EOMphiNLO1NinFrho}), (\ref{eq:EOMFNLO1NinFrho}), and (\ref{eq:EOMrhoNLO1NinFrho}), respectively.
In these equations, the real functions $M_{ij}(x;\phi,F)$, $\Sigma^{(0)}_{ij}(x,y;\phi,F)$, $\Sigma^F_{ij}(x,y;\phi,G)$, and $\Sigma^\rho_{ij}(x,y;\phi,G)$ are given in terms of $\phi$, $F$, and $\rho$ as follows:
\begin{widetext}
\begin{align}
\label{eq:MofF}
  M_{ij}(x,y;\phi,F)
  &= \delta_{ij}\delta_{\cal C}(x-y)
  \Big[H_\mathrm{1B}(x)
   + \frac{1}{2}\int_z V(x-z)\Big(\phi_k(z)\phi_k(z)+F_{kk}(z,z)\Big)\Big]
   + V(x-y)\Big(\phi_i(x)\phi_j(y)+F_{ij}(x,y)\Big),
  \\ 
\label{eq:SigmabarFofFrho}
  \Sigma^F_{ij}(x,y;\phi,G)
  &= -\overline{\Lambda}_F(x,y)\Big[\phi_i(x)\phi_j(y)+F_{ij}(x,y)\Big]
     +\frac{1}{4}\overline{\Lambda}_\rho(x,y)\,\rho_{ij}(x,y)
     -P_F(x,y)\,F_{ij}(x,y)+\frac{1}{4}P_\rho(x,y))\,\rho_{ij}(x,y),
  \\ 
\label{eq:SigmabarrhoofFrho}
  \Sigma^\rho_{ij}(x,y;\phi,G)
  &= -\overline{\Lambda}_\rho(x,y)\Big[\phi_i(x)\phi_j(y)+F_{ij}(x,y)\Big]
     -\overline{\Lambda}_F(x,y)\,\rho_{ij}(x,y)
     -P_\rho(x,y)\,F_{ij}(x,y)-P_F(x,y))\,\rho_{ij}(x,y).
\end{align} 
In here, the functions $\overline{\Lambda}_{F,\rho}(x,y)$ are given as
\begin{align}
\label{eq:LambdabarFrho}
  \overline{\Lambda}_{F,\rho}(x,y)
  &= \frac{2}{\cN}\int_z I_{F,\rho}(x,z)\,V(z-y),
  \\
\label{eq:IF}
  I_F(x,y)
  &= 
    \int_z V(x-z)
    \Big(F(z,y)^2-\frac{1}{4}\rho(z,y)^2\Big)
    -\int_u\Big[\int_0^{x_0}dz\,I_\rho(x,u)\,V(u-z)\,
                 \Big(F(z,y)^2-\frac{1}{4}\rho(z,y)^2\Big)
  \nonumber\\
  &\qquad\qquad
	       -2\int_0^{y_0}dz\,I_F(x,u)\,V(u-z)\,
	         F_{ij}(z,y)\rho_{ij}(z,y)\Big],   
  \\
\label{eq:Irho}
  I_\rho(x,y)
  &= {2}\Big\{
    \int_z V(x-z)
    F_{ij}(z,y)\rho_{ij}(z,y)
    -\int_u\int_{y_0}^{x_0}dz\,I_\rho(x,u)\,V(u-z)\,
                 F_{ij}(z,y)\rho_{ij}(z,y)\Big\},
\end{align} 
where $F^2\equiv F_{ij}F_{ij}$, etc., and the functions $P_{F,\rho}(x,y)$ as
\begin{align}
\label{eq:PF}
  P_F(x,y)
  &= \frac{4}{\cN^2}\int_v\Big\{\int_w V(x-v)\,H_F(v,w)\,V(w-y)
  \nonumber\\
  &\qquad\quad
   +\int_0^{y_0}dw\,\Big[
          V(x-v)\,H_F   (v,w)\,I_\rho(w,y)
    +I_F   (x,w)\,H_\rho(w,v)\,     V(v-y)\Big]
  \nonumber\\
  &\qquad\quad
   -\int_0^{x_0}dw\,\Big[
          V(x-v)\,H_\rho(v,w)\,   I_F(w,y)
    +I_\rho(x,w)\,H_F   (w,v)\,     V(v-y)\Big]
	   \Big\}
  \nonumber\\
  &\qquad
   -\int_0^{x_0}dv\int_0^{y_0}dw\,
     I_\rho(x,v)\,   H_F(v,w)\,I_\rho(w,y)
   +\int_0^{x_0}dv\int_0^{v_0}dw\,
     I_\rho(x,v)\,H_\rho(v,w)\,   I_F(w,y)
  \nonumber\\
  &\qquad
   +\int_0^{y_0}dv\int_{z_0}^{y_0}dw\,
     I_F   (x,v)\,H_\rho(v,w)\,I_\rho(w,y),
  \\
\label{eq:Prho}
  P_\rho(x,y)
  &= \frac{4}{\cN^2}\int_v\Big\{\int_w V(x-v)\,H_\rho(v,w)\,V(w-y)
  \nonumber\\
  &\qquad\quad
   -\int_{y_0}^{x_0}dw\,\Big[
          V(x-v)\,H_\rho(v,w)\,I_\rho(w,y)
    +I_\rho(x,w)\,H_\rho(w,v)\,     V(v-y)\Big]\Big\}
  \nonumber\\
  &\qquad
   +\int_{y_0}^{x_0}dv\int_{y_0}^{v_0}dw\,
     I_\rho(x,v)\,H_\rho(v,w)\,I_\rho(w,y),
\end{align} 
\end{widetext}
with
\begin{align}
\label{eq:HF}
  H_F(x,y)
  &= \phi_i(x)\,F_{ij}(x,y)\,\phi_j(y),
  \\
\label{eq:Hrho}
  H_\rho(x,y)
  &= \phi_i(x)\,\rho_{ij}(x,y)\,\phi_j(y),
\end{align} 
where it is, as usual, summed over double indices.
In Eqs.~(\ref{eq:PF}) and (\ref{eq:Prho}) it has been used that all time integrals over the closed path $\cal C$, of $H_FI_F$, $I_FH_F$, and $I_FH_\rho I_F$ vanish.
   
Finally, we provide the momentum-space self energies $\Sigma^{F,\rho}_{ij}(t,t';p)$ in 1+1 dimensions which enter the dynamical equations (\ref{eq:EOMFpNLO1NinFrho}), (\ref{eq:EOMrhopNLO1NinFrho}).
From Eqs.~(\ref{eq:SigmabarFofFrho}) to (\ref{eq:Irho}), one obtains, for a homogeneous system, with $\phi_i\equiv0$, and a coupling given in Eq.~(\ref{eq:1Dcoupling}), by Fourier transformation:
\begin{align}
\label{eq:SigmabarFpofFrho}
  \Sigma^F_{ij}(t,t';p;G)
  &= -g_\mathrm{1D}\int_k\Big[I_F(t,t';p-k)F_{ij}(t,t';k)
  \nonumber\\
  &\qquad
                  -\frac{1}{4}I_\rho(t,t';p-k)\,\rho_{ij}(t,t';k)\Big],
  \\ 
\label{eq:SigmabarrhopofFrho}
  \Sigma^\rho_{ij}(t,t';p;G)
  &= -g_\mathrm{1D}\int_k\Big[I_\rho(t,t';p-k)F_{ij}(t,t';k)
  \nonumber\\
  &\qquad
                             +I_F(t,t';p-k)\rho_{ij}(t,t';k)\Big],
\end{align} 
\begin{align}
\label{eq:Irhop}
  &I_\rho(t,t';p)
  = {2}g_\mathrm{1D}\int_{k}\Big\{
    F_{ij}(t,t';p-k)\rho_{ij}(t,t';k)
  \nonumber\\
  &\qquad
    -\int_{k'}\int_{t'}^{t}dt''\,I_\rho(t,t'';p-k)\,
  \nonumber\\
  &\qquad\qquad\qquad\times
                 F_{ij}(t,t';k-k')\rho_{ij}(t,t';k')\Big\},
\end{align} 
\begin{align}
\label{eq:IFp}
  &I_F(t,t';p)
  =g_\mathrm{1D}\int_{k}\Big\{
    F_{ij}(t,t';p-k)F_{ij}(t,t';k)
  \nonumber\\
  &\qquad\qquad\qquad\qquad
    -\frac{1}{4}\rho_{ij}(t,t';p-k)\rho_{ij}(t,t';k)
  \nonumber\\
  &\qquad
    -\int_{k'}\Big[\int_0^tdt''\,I_\rho(t,t'';p-k)\,
  \nonumber\\
  &\qquad\qquad\qquad\times
              \Big(F_{ij}(t'',t';k-k')F_{ij}(t'',t';k')
  \nonumber\\
  &\qquad\qquad\qquad
    -\frac{1}{4}\rho_{ij}(t'',t';k-k')\rho_{ij}(t'',t';k')\Big)
  \nonumber\\
  &\qquad
    +2\int_0^{t'}dt''\,I_F(t,t'';p-k)\,
  \nonumber\\
  &\qquad\qquad\qquad\times
	         F_{ij}(t,t';k-k')\rho_{ij}(t,t';k')\Big]\Big\},   
\end{align} 
Here, $\int_k\equiv(2\pi)^{-1}\int dk$ denotes the one-dimensional momentum integral.

\end{appendix}

\newpage
\bibliography{TGbib}

\begin{thebibliography}{87}
\expandafter\ifx\csname natexlab\endcsname\relax\def\natexlab#1{#1}\fi
\expandafter\ifx\csname bibnamefont\endcsname\relax
  \def\bibnamefont#1{#1}\fi
\expandafter\ifx\csname bibfnamefont\endcsname\relax
  \def\bibfnamefont#1{#1}\fi
\expandafter\ifx\csname citenamefont\endcsname\relax
  \def\citenamefont#1{#1}\fi
\expandafter\ifx\csname url\endcsname\relax
  \def\url#1{\texttt{#1}}\fi
\expandafter\ifx\csname urlprefix\endcsname\relax\def\urlprefix{URL }\fi
\providecommand{\bibinfo}[2]{#2}
\providecommand{\eprint}[2][]{\url{#2}}

\bibitem[{\citenamefont{Goss-Levi}(2000)}]{GossLevi2000a}
\bibinfo{author}{\bibfnamefont{B.}~\bibnamefont{Goss-Levi}},
  \bibinfo{journal}{Phys. Today} \textbf{\bibinfo{volume}{53}},
  \bibinfo{pages}{46} (\bibinfo{year}{2000}).

\bibitem[{\citenamefont{Williams and Julienne}(2000)}]{Williams2000a}
\bibinfo{author}{\bibfnamefont{C.~J.} \bibnamefont{Williams}} \bibnamefont{and}
  \bibinfo{author}{\bibfnamefont{P.~S.} \bibnamefont{Julienne}},
  \bibinfo{journal}{Science} \textbf{\bibinfo{volume}{287}},
  \bibinfo{pages}{986} (\bibinfo{year}{2000}).

\bibitem[{\citenamefont{Weiner et~al.}(1999)\citenamefont{Weiner, Bagnato,
  Zilio, and Julienne}}]{Weiner1999a}
\bibinfo{author}{\bibfnamefont{J.}~\bibnamefont{Weiner}},
  \bibinfo{author}{\bibfnamefont{V.~S.} \bibnamefont{Bagnato}},
  \bibinfo{author}{\bibfnamefont{S.}~\bibnamefont{Zilio}}, \bibnamefont{and}
  \bibinfo{author}{\bibfnamefont{P.~S.} \bibnamefont{Julienne}},
  \bibinfo{journal}{Rev. Mod. Phys.} \textbf{\bibinfo{volume}{71}},
  \bibinfo{pages}{1} (\bibinfo{year}{1999}).

\bibitem[{\citenamefont{Heinzen et~al.}(2000)\citenamefont{Heinzen, Wynar,
  Drummond, and Kheruntsyan}}]{Heinzen2000a}
\bibinfo{author}{\bibfnamefont{D.~J.} \bibnamefont{Heinzen}},
  \bibinfo{author}{\bibfnamefont{R.}~\bibnamefont{Wynar}},
  \bibinfo{author}{\bibfnamefont{P.~D.} \bibnamefont{Drummond}},
  \bibnamefont{and} \bibinfo{author}{\bibfnamefont{K.~V.}
  \bibnamefont{Kheruntsyan}}, \bibinfo{journal}{Phys. Rev. Lett.}
  \textbf{\bibinfo{volume}{84}}, \bibinfo{pages}{5029} (\bibinfo{year}{2000}).

\bibitem[{\citenamefont{Randeria}(1995)}]{Randeria1995a}
\bibinfo{author}{\bibfnamefont{M.}~\bibnamefont{Randeria}}, in
  \emph{\bibinfo{booktitle}{Bose-Einstein condensation}}, edited by
  \bibinfo{editor}{\bibfnamefont{A.}~\bibnamefont{Griffin}},
  \bibinfo{editor}{\bibfnamefont{D.~W.} \bibnamefont{Snoke}}, \bibnamefont{and}
  \bibinfo{editor}{\bibfnamefont{S.}~\bibnamefont{Stringari}}
  (\bibinfo{publisher}{Cambridge University Press, Cambridge, UK},
  \bibinfo{year}{1995}), vol. \bibinfo{volume}{392}, p. \bibinfo{pages}{355}.

\bibitem[{\citenamefont{Schmiedmayer and Folman}(2001)}]{Schmiedmayer2000a}
\bibinfo{author}{\bibfnamefont{J.}~\bibnamefont{Schmiedmayer}}
  \bibnamefont{and} \bibinfo{author}{\bibfnamefont{R.}~\bibnamefont{Folman}},
  \bibinfo{journal}{C. R. Acad. Sci. Paris IV} \textbf{\bibinfo{volume}{2}},
  \bibinfo{pages}{333} (\bibinfo{year}{2001}).

\bibitem[{\citenamefont{Pitaevskii and Stringari}(2003)}]{Pitaevskii2003a}
\bibinfo{author}{\bibfnamefont{L.~P.} \bibnamefont{Pitaevskii}}
  \bibnamefont{and}
  \bibinfo{author}{\bibfnamefont{S.}~\bibnamefont{Stringari}},
  \emph{\bibinfo{title}{{B}ose-{E}instein Condensation}}
  (\bibinfo{publisher}{Clarendon Press}, \bibinfo{year}{2003}).

\bibitem[{\citenamefont{Jaksch et~al.}(1998)\citenamefont{Jaksch, Bruder,
  Cirac, Gardiner, and Zoller}}]{Jaksch1998a}
\bibinfo{author}{\bibfnamefont{D.}~\bibnamefont{Jaksch}},
  \bibinfo{author}{\bibnamefont{Bruder}}, \bibinfo{author}{\bibfnamefont{J.~I.}
  \bibnamefont{Cirac}}, \bibinfo{author}{\bibfnamefont{C.~W.}
  \bibnamefont{Gardiner}}, \bibnamefont{and}
  \bibinfo{author}{\bibfnamefont{P.}~\bibnamefont{Zoller}},
  \bibinfo{journal}{Phys. Rev. Lett.} \textbf{\bibinfo{volume}{81}},
  \bibinfo{pages}{3108} (\bibinfo{year}{1998}).

\bibitem[{\citenamefont{Bloch}(2004)}]{Bloch2004a}
\bibinfo{author}{\bibfnamefont{I.}~\bibnamefont{Bloch}},
  \bibinfo{journal}{Phys. World} \textbf{\bibinfo{volume}{17}},
  \bibinfo{pages}{25} (\bibinfo{year}{2004}).

\bibitem[{\citenamefont{Stwalley}(1976)}]{Stwalley1976b}
\bibinfo{author}{\bibfnamefont{W.~C.} \bibnamefont{Stwalley}},
  \bibinfo{journal}{Phys. Rev. Lett.} \textbf{\bibinfo{volume}{37}},
  \bibinfo{pages}{1628} (\bibinfo{year}{1976}).

\bibitem[{\citenamefont{Tiesinga et~al.}(1992)\citenamefont{Tiesinga, Moerdijk,
  Verhaar, and Stoof}}]{Tiesinga1992a}
\bibinfo{author}{\bibfnamefont{E.}~\bibnamefont{Tiesinga}},
  \bibinfo{author}{\bibfnamefont{A.}~\bibnamefont{Moerdijk}},
  \bibinfo{author}{\bibfnamefont{B.~J.} \bibnamefont{Verhaar}},
  \bibnamefont{and} \bibinfo{author}{\bibfnamefont{H.~T.~C.}
  \bibnamefont{Stoof}}, \bibinfo{journal}{Phys. Rev. A}
  \textbf{\bibinfo{volume}{46}}, \bibinfo{pages}{R1167} (\bibinfo{year}{1992}).

\bibitem[{\citenamefont{Tiesinga et~al.}(1993)\citenamefont{Tiesinga, Verhaar,
  and Stoof}}]{Tiesinga1993a}
\bibinfo{author}{\bibfnamefont{E.}~\bibnamefont{Tiesinga}},
  \bibinfo{author}{\bibfnamefont{B.~J.} \bibnamefont{Verhaar}},
  \bibnamefont{and} \bibinfo{author}{\bibfnamefont{H.~T.~C.}
  \bibnamefont{Stoof}}, \bibinfo{journal}{Phys. Rev. A}
  \textbf{\bibinfo{volume}{47}}, \bibinfo{pages}{4114} (\bibinfo{year}{1993}).

\bibitem[{\citenamefont{Burnett}(1998)}]{Burnett1998a}
\bibinfo{author}{\bibfnamefont{K.}~\bibnamefont{Burnett}},
  \bibinfo{journal}{Nature (London)} \textbf{\bibinfo{volume}{392}},
  \bibinfo{pages}{125} (\bibinfo{year}{1998}).

\bibitem[{\citenamefont{Mies et~al.}(2000)\citenamefont{Mies, Tiesinga, and
  Julienne}}]{Mies2000a}
\bibinfo{author}{\bibfnamefont{F.~H.} \bibnamefont{Mies}},
  \bibinfo{author}{\bibfnamefont{E.}~\bibnamefont{Tiesinga}}, \bibnamefont{and}
  \bibinfo{author}{\bibfnamefont{P.~S.} \bibnamefont{Julienne}},
  \bibinfo{journal}{Phys. Rev. A} \textbf{\bibinfo{volume}{61}},
  \bibinfo{pages}{022721} (\bibinfo{year}{2000}).

\bibitem[{\citenamefont{Regal et~al.}(2004)\citenamefont{Regal, Greiner, and
  Jin}}]{Regal2004b}
\bibinfo{author}{\bibfnamefont{C.~A.} \bibnamefont{Regal}},
  \bibinfo{author}{\bibfnamefont{M.}~\bibnamefont{Greiner}}, \bibnamefont{and}
  \bibinfo{author}{\bibfnamefont{D.~S.} \bibnamefont{Jin}},
  \bibinfo{journal}{Phys. Rev. Lett.} \textbf{\bibinfo{volume}{92}},
  \bibinfo{pages}{040403} (\bibinfo{year}{2004}).

\bibitem[{\citenamefont{Zwierlein et~al.}(2004)\citenamefont{Zwierlein, Stan,
  Schunck, Raupach, Kerman, and Ketterle}}]{Zwierlein2004a}
\bibinfo{author}{\bibfnamefont{M.~W.} \bibnamefont{Zwierlein}},
  \bibinfo{author}{\bibfnamefont{C.~A.} \bibnamefont{Stan}},
  \bibinfo{author}{\bibfnamefont{C.~H.} \bibnamefont{Schunck}},
  \bibinfo{author}{\bibfnamefont{S.~M.~F.} \bibnamefont{Raupach}},
  \bibinfo{author}{\bibfnamefont{A.~J.} \bibnamefont{Kerman}},
  \bibnamefont{and} \bibinfo{author}{\bibfnamefont{W.}~\bibnamefont{Ketterle}},
  \bibinfo{journal}{Phys. Rev. Lett.} \textbf{\bibinfo{volume}{92}},
  \bibinfo{pages}{120403} (\bibinfo{year}{2004}).

\bibitem[{\citenamefont{Bartenstein et~al.}(2004)\citenamefont{Bartenstein,
  Altmeyer, Riedl, Jochim, Chin, Denschlag, and Grimm}}]{Bartenstein2004a}
\bibinfo{author}{\bibfnamefont{M.}~\bibnamefont{Bartenstein}},
  \bibinfo{author}{\bibfnamefont{A.}~\bibnamefont{Altmeyer}},
  \bibinfo{author}{\bibfnamefont{S.}~\bibnamefont{Riedl}},
  \bibinfo{author}{\bibfnamefont{S.}~\bibnamefont{Jochim}},
  \bibinfo{author}{\bibfnamefont{C.}~\bibnamefont{Chin}},
  \bibinfo{author}{\bibfnamefont{J.~H.} \bibnamefont{Denschlag}},
  \bibnamefont{and} \bibinfo{author}{\bibfnamefont{R.}~\bibnamefont{Grimm}},
  \bibinfo{journal}{Phys. Rev. Lett.} \textbf{\bibinfo{volume}{92}},
  \bibinfo{pages}{120401} (\bibinfo{year}{2004}).

\bibitem[{\citenamefont{Tonks}(1936)}]{Tonks1936a}
\bibinfo{author}{\bibfnamefont{L.}~\bibnamefont{Tonks}},
  \bibinfo{journal}{Phys. Rev.} \textbf{\bibinfo{volume}{50}},
  \bibinfo{pages}{955} (\bibinfo{year}{1936}).

\bibitem[{\citenamefont{Girardeau}(1960)}]{Girardeau1960a}
\bibinfo{author}{\bibfnamefont{M.}~\bibnamefont{Girardeau}},
  \bibinfo{journal}{J. Math. Phys. (NY)} \textbf{\bibinfo{volume}{1}},
  \bibinfo{pages}{516} (\bibinfo{year}{1960}).

\bibitem[{\citenamefont{Paredes et~al.}(2004)\citenamefont{Paredes, Widera,
  Murg, Mandel, F{\"o}lling, Cirac, Shlyapnikov, H{\"a}nsch, and
  Bloch}}]{Paredes2004a}
\bibinfo{author}{\bibfnamefont{B.}~\bibnamefont{Paredes}},
  \bibinfo{author}{\bibfnamefont{A.}~\bibnamefont{Widera}},
  \bibinfo{author}{\bibfnamefont{V.}~\bibnamefont{Murg}},
  \bibinfo{author}{\bibfnamefont{O.}~\bibnamefont{Mandel}},
  \bibinfo{author}{\bibfnamefont{S.}~\bibnamefont{F{\"o}lling}},
  \bibinfo{author}{\bibfnamefont{I.}~\bibnamefont{Cirac}},
  \bibinfo{author}{\bibfnamefont{G.~V.} \bibnamefont{Shlyapnikov}},
  \bibinfo{author}{\bibfnamefont{T.~W.} \bibnamefont{H{\"a}nsch}},
  \bibnamefont{and} \bibinfo{author}{\bibfnamefont{I.}~\bibnamefont{Bloch}},
  \bibinfo{journal}{Nature (London)} \textbf{\bibinfo{volume}{429}},
  \bibinfo{pages}{277} (\bibinfo{year}{2004}).

\bibitem[{\citenamefont{van Oosten et~al.}(2001)\citenamefont{van Oosten,
  van~der Straten, and Stoof}}]{vanOosten2001a}
\bibinfo{author}{\bibfnamefont{D.}~\bibnamefont{van Oosten}},
  \bibinfo{author}{\bibfnamefont{P.}~\bibnamefont{van~der Straten}},
  \bibnamefont{and} \bibinfo{author}{\bibfnamefont{H.~T.~C.}
  \bibnamefont{Stoof}}, \bibinfo{journal}{Phys. Rev. A}
  \textbf{\bibinfo{volume}{63}}, \bibinfo{pages}{053601}
  (\bibinfo{year}{2001}).

\bibitem[{\citenamefont{Greiner et~al.}(2002)\citenamefont{Greiner, Mandel,
  Esslinger, H{\"a}nsch, and Bloch}}]{Greiner2002a}
\bibinfo{author}{\bibfnamefont{M.}~\bibnamefont{Greiner}},
  \bibinfo{author}{\bibfnamefont{O.}~\bibnamefont{Mandel}},
  \bibinfo{author}{\bibfnamefont{T.}~\bibnamefont{Esslinger}},
  \bibinfo{author}{\bibfnamefont{T.~W.} \bibnamefont{H{\"a}nsch}},
  \bibnamefont{and} \bibinfo{author}{\bibfnamefont{I.}~\bibnamefont{Bloch}},
  \bibinfo{journal}{Nature (London)} \textbf{\bibinfo{volume}{415}},
  \bibinfo{pages}{39} (\bibinfo{year}{2002}).

\bibitem[{\citenamefont{Roberts et~al.}(2001)\citenamefont{Roberts, Claussen,
  Cornish, Donley, Cornell, and Wieman}}]{Roberts2001a}
\bibinfo{author}{\bibfnamefont{J.~L.} \bibnamefont{Roberts}},
  \bibinfo{author}{\bibfnamefont{N.~R.} \bibnamefont{Claussen}},
  \bibinfo{author}{\bibfnamefont{S.~L.} \bibnamefont{Cornish}},
  \bibinfo{author}{\bibfnamefont{E.~A.} \bibnamefont{Donley}},
  \bibinfo{author}{\bibfnamefont{E.~A.} \bibnamefont{Cornell}},
  \bibnamefont{and} \bibinfo{author}{\bibfnamefont{C.~E.}
  \bibnamefont{Wieman}}, \bibinfo{journal}{Phys. Rev. Lett.}
  \textbf{\bibinfo{volume}{86}}, \bibinfo{pages}{4211} (\bibinfo{year}{2001}).

\bibitem[{\citenamefont{Donley et~al.}(2002)\citenamefont{Donley, Claussen,
  Thompson, and Wieman}}]{Donley2002a}
\bibinfo{author}{\bibfnamefont{E.~A.} \bibnamefont{Donley}},
  \bibinfo{author}{\bibfnamefont{N.~R.} \bibnamefont{Claussen}},
  \bibinfo{author}{\bibfnamefont{S.~T.} \bibnamefont{Thompson}},
  \bibnamefont{and} \bibinfo{author}{\bibfnamefont{C.~E.}
  \bibnamefont{Wieman}}, \bibinfo{journal}{Nature (London)}
  \textbf{\bibinfo{volume}{417}}, \bibinfo{pages}{529} (\bibinfo{year}{2002}).

\bibitem[{\citenamefont{Kagan et~al.}(1992)\citenamefont{Kagan, Svistunov, and
  Shlyapnikov}}]{Kagan1992a}
\bibinfo{author}{\bibfnamefont{Y.}~\bibnamefont{Kagan}},
  \bibinfo{author}{\bibfnamefont{B.~V.} \bibnamefont{Svistunov}},
  \bibnamefont{and} \bibinfo{author}{\bibfnamefont{G.~V.}
  \bibnamefont{Shlyapnikov}}, \bibinfo{journal}{[Zh. Eksp. Teor. Fiz. 101, 528
  (1992)] Sov. Phys. JETP} \textbf{\bibinfo{volume}{75}}, \bibinfo{pages}{387}
  (\bibinfo{year}{1992}).

\bibitem[{\citenamefont{Stoof}(1991)}]{Stoof1991a}
\bibinfo{author}{\bibfnamefont{H.~T.~C.} \bibnamefont{Stoof}},
  \bibinfo{journal}{Phys. Rev. Lett.} \textbf{\bibinfo{volume}{66}},
  \bibinfo{pages}{3148} (\bibinfo{year}{1991}).

\bibitem[{\citenamefont{Stoof}(1997)}]{Stoof1997a}
\bibinfo{author}{\bibfnamefont{H.~T.~C.} \bibnamefont{Stoof}},
  \bibinfo{journal}{Phys. Rev. Lett.} \textbf{\bibinfo{volume}{78}},
  \bibinfo{pages}{768} (\bibinfo{year}{1997}).

\bibitem[{\citenamefont{Gardiner et~al.}(1998)\citenamefont{Gardiner, Lee,
  Ballagh, Davis, and Zoller}}]{Gardiner1998b}
\bibinfo{author}{\bibfnamefont{C.~W.} \bibnamefont{Gardiner}},
  \bibinfo{author}{\bibfnamefont{M.~D.} \bibnamefont{Lee}},
  \bibinfo{author}{\bibfnamefont{R.~J.} \bibnamefont{Ballagh}},
  \bibinfo{author}{\bibfnamefont{M.~J.} \bibnamefont{Davis}}, \bibnamefont{and}
  \bibinfo{author}{\bibfnamefont{P.}~\bibnamefont{Zoller}},
  \bibinfo{journal}{Phys. Rev. Lett.} \textbf{\bibinfo{volume}{81}},
  \bibinfo{pages}{5266} (\bibinfo{year}{1998}).

\bibitem[{\citenamefont{Davis and Gardiner}(2002)}]{Davis2002a}
\bibinfo{author}{\bibfnamefont{M.~J.} \bibnamefont{Davis}} \bibnamefont{and}
  \bibinfo{author}{\bibfnamefont{C.~W.} \bibnamefont{Gardiner}},
  \bibinfo{journal}{J. Phys. B} \textbf{\bibinfo{volume}{35}},
  \bibinfo{pages}{733} (\bibinfo{year}{2002}).

\bibitem[{\citenamefont{Davis et~al.}(2002)\citenamefont{Davis, Morgan, and
  Burnett}}]{Davis2002b}
\bibinfo{author}{\bibfnamefont{M.~J.} \bibnamefont{Davis}},
  \bibinfo{author}{\bibfnamefont{S.~A.} \bibnamefont{Morgan}},
  \bibnamefont{and} \bibinfo{author}{\bibfnamefont{K.}~\bibnamefont{Burnett}},
  \bibinfo{journal}{Phys. Rev. A} \textbf{\bibinfo{volume}{66}},
  \bibinfo{pages}{053618} (\bibinfo{year}{2002}).

\bibitem[{\citenamefont{K{\"o}hl et~al.}(2002)\citenamefont{K{\"o}hl, Davis,
  Gardiner, H{\"a}nsch, and Esslinger}}]{Kohl2002a}
\bibinfo{author}{\bibfnamefont{M.}~\bibnamefont{K{\"o}hl}},
  \bibinfo{author}{\bibfnamefont{M.~J.} \bibnamefont{Davis}},
  \bibinfo{author}{\bibfnamefont{C.~W.} \bibnamefont{Gardiner}},
  \bibinfo{author}{\bibfnamefont{T.}~\bibnamefont{H{\"a}nsch}},
  \bibnamefont{and}
  \bibinfo{author}{\bibfnamefont{T.}~\bibnamefont{Esslinger}},
  \bibinfo{journal}{Phys. Rev. Lett.} \textbf{\bibinfo{volume}{88}},
  \bibinfo{pages}{080402} (\bibinfo{year}{2002}).

\bibitem[{\citenamefont{Cornwall et~al.}(1974)\citenamefont{Cornwall, Jackiw,
  and Tomboulis}}]{Cornwall1974a}
\bibinfo{author}{\bibfnamefont{J.~M.} \bibnamefont{Cornwall}},
  \bibinfo{author}{\bibfnamefont{R.}~\bibnamefont{Jackiw}}, \bibnamefont{and}
  \bibinfo{author}{\bibfnamefont{E.}~\bibnamefont{Tomboulis}},
  \bibinfo{journal}{Phys. Rev. D} \textbf{\bibinfo{volume}{10}},
  \bibinfo{pages}{2428} (\bibinfo{year}{1974}).

\bibitem[{\citenamefont{Berges}(2002)}]{Berges2002a}
\bibinfo{author}{\bibfnamefont{J.}~\bibnamefont{Berges}},
  \bibinfo{journal}{Nucl. Phys.} \textbf{\bibinfo{volume}{A699}},
  \bibinfo{pages}{847} (\bibinfo{year}{2002}).

\bibitem[{\citenamefont{Aarts et~al.}(2002)\citenamefont{Aarts, Ahrensmeier,
  Baier, Berges, and Serreau}}]{Aarts2002b}
\bibinfo{author}{\bibfnamefont{G.}~\bibnamefont{Aarts}},
  \bibinfo{author}{\bibfnamefont{D.}~\bibnamefont{Ahrensmeier}},
  \bibinfo{author}{\bibfnamefont{R.}~\bibnamefont{Baier}},
  \bibinfo{author}{\bibfnamefont{J.}~\bibnamefont{Berges}}, \bibnamefont{and}
  \bibinfo{author}{\bibfnamefont{J.}~\bibnamefont{Serreau}},
  \bibinfo{journal}{Phys. Rev. D} \textbf{\bibinfo{volume}{66}},
  \bibinfo{pages}{045008} (\bibinfo{year}{2002}).

\bibitem[{\citenamefont{Berges and Serreau}(2003)}]{Berges2003b}
\bibinfo{author}{\bibfnamefont{J.}~\bibnamefont{Berges}} \bibnamefont{and}
  \bibinfo{author}{\bibfnamefont{J.}~\bibnamefont{Serreau}},
  \bibinfo{journal}{Phys. Rev. Lett.} \textbf{\bibinfo{volume}{91}},
  \bibinfo{pages}{111601} (\bibinfo{year}{2003}).

\bibitem[{\citenamefont{Mihaila et~al.}(2001)\citenamefont{Mihaila, Dawson, and
  Cooper}}]{Mihaila2001a}
\bibinfo{author}{\bibfnamefont{B.}~\bibnamefont{Mihaila}},
  \bibinfo{author}{\bibfnamefont{J.~F.} \bibnamefont{Dawson}},
  \bibnamefont{and} \bibinfo{author}{\bibfnamefont{F.}~\bibnamefont{Cooper}},
  \bibinfo{journal}{Phys. Rev. D} \textbf{\bibinfo{volume}{63}},
  \bibinfo{pages}{096003} (\bibinfo{year}{2001}).

\bibitem[{\citenamefont{Cooper et~al.}(2003)\citenamefont{Cooper, Dawson, and
  Mihaila}}]{Cooper2003a}
\bibinfo{author}{\bibfnamefont{F.}~\bibnamefont{Cooper}},
  \bibinfo{author}{\bibfnamefont{J.~F.} \bibnamefont{Dawson}},
  \bibnamefont{and} \bibinfo{author}{\bibfnamefont{B.}~\bibnamefont{Mihaila}},
  \bibinfo{journal}{Phys. Rev. D} \textbf{\bibinfo{volume}{67}},
  \bibinfo{pages}{056003} (\bibinfo{year}{2003}).

\bibitem[{\citenamefont{Arrizabalaga et~al.}(2004)\citenamefont{Arrizabalaga,
  Smit, and Tranberg}}]{Arrizabalaga2004a}
\bibinfo{author}{\bibfnamefont{A.}~\bibnamefont{Arrizabalaga}},
  \bibinfo{author}{\bibfnamefont{J.}~\bibnamefont{Smit}}, \bibnamefont{and}
  \bibinfo{author}{\bibfnamefont{A.}~\bibnamefont{Tranberg}},
  \bibinfo{journal}{JHEP} \textbf{\bibinfo{volume}{10}}, \bibinfo{pages}{017}
  (\bibinfo{year}{2004}).

\bibitem[{\citenamefont{Berges et~al.}(2003)\citenamefont{Berges,
  Bors{\'{a}}nyi, and Serreau}}]{Berges2003a}
\bibinfo{author}{\bibfnamefont{J.}~\bibnamefont{Berges}},
  \bibinfo{author}{\bibfnamefont{S.}~\bibnamefont{Bors{\'{a}}nyi}},
  \bibnamefont{and} \bibinfo{author}{\bibfnamefont{J.}~\bibnamefont{Serreau}},
  \bibinfo{journal}{Nucl. Phys.} \textbf{\bibinfo{volume}{B660}},
  \bibinfo{pages}{51} (\bibinfo{year}{2003}).

\bibitem[{\citenamefont{Berges et~al.}(2004)\citenamefont{Berges,
  Bors{\'{a}}nyi, and Wetterich}}]{Berges2004b}
\bibinfo{author}{\bibfnamefont{J.}~\bibnamefont{Berges}},
  \bibinfo{author}{\bibfnamefont{S.}~\bibnamefont{Bors{\'{a}}nyi}},
  \bibnamefont{and}
  \bibinfo{author}{\bibfnamefont{C.}~\bibnamefont{Wetterich}},
  \bibinfo{journal}{Phys. Rev. Lett.} \textbf{\bibinfo{volume}{93}},
  \bibinfo{pages}{142002} (\bibinfo{year}{2004}).

\bibitem[{\citenamefont{Rey et~al.}(2004)\citenamefont{Rey, Hu, Calzetta,
  Roura, and Clark}}]{Rey2004a}
\bibinfo{author}{\bibfnamefont{A.}~\bibnamefont{Rey}},
  \bibinfo{author}{\bibfnamefont{B.}~\bibnamefont{Hu}},
  \bibinfo{author}{\bibfnamefont{E.}~\bibnamefont{Calzetta}},
  \bibinfo{author}{\bibfnamefont{A.}~\bibnamefont{Roura}}, \bibnamefont{and}
  \bibinfo{author}{\bibfnamefont{C.}~\bibnamefont{Clark}},
  \bibinfo{journal}{Phys. Rev. A} \textbf{\bibinfo{volume}{69}},
  \bibinfo{pages}{033610} (\bibinfo{year}{2004}).

\bibitem[{\citenamefont{Andersen}(2004)}]{Andersen2004a}
\bibinfo{author}{\bibfnamefont{J.~O.} \bibnamefont{Andersen}},
  \bibinfo{journal}{Rev. Mod. Phys.} \textbf{\bibinfo{volume}{76}},
  \bibinfo{pages}{599} (\bibinfo{year}{2004}).

\bibitem[{\citenamefont{Proukakis and Burnett}(1996)}]{Proukakis1996a}
\bibinfo{author}{\bibfnamefont{N.~P.} \bibnamefont{Proukakis}}
  \bibnamefont{and} \bibinfo{author}{\bibfnamefont{K.}~\bibnamefont{Burnett}},
  \bibinfo{journal}{J. Res. Natl. Inst. Stand. Technol.}
  \textbf{\bibinfo{volume}{101}}, \bibinfo{pages}{457} (\bibinfo{year}{1996}).

\bibitem[{\citenamefont{Shi and Griffin}(1998)}]{Shi1998a}
\bibinfo{author}{\bibfnamefont{H.}~\bibnamefont{Shi}} \bibnamefont{and}
  \bibinfo{author}{\bibfnamefont{A.}~\bibnamefont{Griffin}},
  \bibinfo{journal}{Phys. Rep.} \textbf{\bibinfo{volume}{304}},
  \bibinfo{pages}{1} (\bibinfo{year}{1998}).

\bibitem[{\citenamefont{Giorgini}(1998)}]{Giorgini1998a}
\bibinfo{author}{\bibfnamefont{S.}~\bibnamefont{Giorgini}},
  \bibinfo{journal}{Phys. Rev. A} \textbf{\bibinfo{volume}{57}},
  \bibinfo{pages}{2949} (\bibinfo{year}{1998}).

\bibitem[{\citenamefont{Walser et~al.}(1999)\citenamefont{Walser, Williams,
  Cooper, and Holland}}]{Walser1999a}
\bibinfo{author}{\bibfnamefont{R.}~\bibnamefont{Walser}},
  \bibinfo{author}{\bibfnamefont{J.}~\bibnamefont{Williams}},
  \bibinfo{author}{\bibfnamefont{J.}~\bibnamefont{Cooper}}, \bibnamefont{and}
  \bibinfo{author}{\bibfnamefont{M.}~\bibnamefont{Holland}},
  \bibinfo{journal}{Phys. Rev. A} \textbf{\bibinfo{volume}{59}},
  \bibinfo{pages}{3878} (\bibinfo{year}{1999}).

\bibitem[{\citenamefont{Bhongale et~al.}(2002)\citenamefont{Bhongale, Walser,
  and Holland}}]{Bhongale2002a}
\bibinfo{author}{\bibfnamefont{S.~G.} \bibnamefont{Bhongale}},
  \bibinfo{author}{\bibfnamefont{R.}~\bibnamefont{Walser}}, \bibnamefont{and}
  \bibinfo{author}{\bibfnamefont{M.~J.} \bibnamefont{Holland}},
  \bibinfo{journal}{Phys. Rev. A} \textbf{\bibinfo{volume}{66}},
  \bibinfo{pages}{043618} (\bibinfo{year}{2002}).

\bibitem[{\citenamefont{Davis et~al.}(2000)\citenamefont{Davis, Gardiner, and
  Ballagh}}]{Davis2000a}
\bibinfo{author}{\bibfnamefont{M.~J.} \bibnamefont{Davis}},
  \bibinfo{author}{\bibfnamefont{C.~W.} \bibnamefont{Gardiner}},
  \bibnamefont{and} \bibinfo{author}{\bibfnamefont{R.~J.}
  \bibnamefont{Ballagh}}, \bibinfo{journal}{Phys. Rev. A}
  \textbf{\bibinfo{volume}{62}}, \bibinfo{pages}{063608}
  (\bibinfo{year}{2000}).

\bibitem[{\citenamefont{Stoof}(1999)}]{Stoof1999a}
\bibinfo{author}{\bibfnamefont{H.~T.~C.} \bibnamefont{Stoof}},
  \bibinfo{journal}{J. Low Temp. Phys.} \textbf{\bibinfo{volume}{114}},
  \bibinfo{pages}{11} (\bibinfo{year}{1999}).

\bibitem[{\citenamefont{Imamovi{\'c}-Tomasovi{\'c} and
  Griffin}(2001)}]{Imamovic-Tomasovic2001a}
\bibinfo{author}{\bibfnamefont{M.}~\bibnamefont{Imamovi{\'c}-Tomasovi{\'c}}}
  \bibnamefont{and} \bibinfo{author}{\bibfnamefont{A.}~\bibnamefont{Griffin}},
  \bibinfo{journal}{J. Low Temp. Phys.} \textbf{\bibinfo{volume}{122}},
  \bibinfo{pages}{617} (\bibinfo{year}{2001}).

\bibitem[{\citenamefont{Boyanovsky et~al.}(2002)\citenamefont{Boyanovsky, Wang,
  Lee, Yu, and Alamoudi}}]{Boyanovsky2002a}
\bibinfo{author}{\bibfnamefont{D.}~\bibnamefont{Boyanovsky}},
  \bibinfo{author}{\bibfnamefont{S.-Y.} \bibnamefont{Wang}},
  \bibinfo{author}{\bibfnamefont{D.-S.} \bibnamefont{Lee}},
  \bibinfo{author}{\bibfnamefont{H.-L.} \bibnamefont{Yu}}, \bibnamefont{and}
  \bibinfo{author}{\bibfnamefont{S.~M.} \bibnamefont{Alamoudi}},
  \bibinfo{journal}{Ann. Phys. (NY)} \textbf{\bibinfo{volume}{300}},
  \bibinfo{pages}{1} (\bibinfo{year}{2002}).

\bibitem[{\citenamefont{Rey et~al.}(2005)\citenamefont{Rey, Hu, Calzetta, and
  Clark}}]{Rey2005a}
\bibinfo{author}{\bibfnamefont{A.~M.} \bibnamefont{Rey}},
  \bibinfo{author}{\bibfnamefont{B.~L.} \bibnamefont{Hu}},
  \bibinfo{author}{\bibfnamefont{E.}~\bibnamefont{Calzetta}}, \bibnamefont{and}
  \bibinfo{author}{\bibfnamefont{C.~W.} \bibnamefont{Clark}},
  \bibinfo{journal}{Phys. Rev. A} \textbf{\bibinfo{volume}{72}},
  \bibinfo{pages}{023604} (\bibinfo{year}{2005}).

\bibitem[{\citenamefont{Baier and Stockamp}(2004)}]{Baier2005a}
\bibinfo{author}{\bibfnamefont{R.}~\bibnamefont{Baier}} \bibnamefont{and}
  \bibinfo{author}{\bibfnamefont{T.}~\bibnamefont{Stockamp}},
  \bibinfo{journal}{eprint hep-ph/0412310}  (\bibinfo{year}{2004}).

\bibitem[{\citenamefont{Calzetta and Hu}(1988)}]{Calzetta1988a}
\bibinfo{author}{\bibfnamefont{E.~A.} \bibnamefont{Calzetta}} \bibnamefont{and}
  \bibinfo{author}{\bibfnamefont{B.~L.} \bibnamefont{Hu}},
  \bibinfo{journal}{Phys. Rev. D} \textbf{\bibinfo{volume}{37}},
  \bibinfo{pages}{2878} (\bibinfo{year}{1988}).

\bibitem[{\citenamefont{Ivanov et~al.}(1999)\citenamefont{Ivanov, Knoll, and
  Voskresensky}}]{Ivanov1999a}
\bibinfo{author}{\bibfnamefont{Y.~B.} \bibnamefont{Ivanov}},
  \bibinfo{author}{\bibfnamefont{J.}~\bibnamefont{Knoll}}, \bibnamefont{and}
  \bibinfo{author}{\bibfnamefont{D.~N.} \bibnamefont{Voskresensky}},
  \bibinfo{journal}{Nucl. Phys.} \textbf{\bibinfo{volume}{A657}},
  \bibinfo{pages}{413} (\bibinfo{year}{1999}).

\bibitem[{\citenamefont{Prokopec
  et~al.}(2004{\natexlab{a}})\citenamefont{Prokopec, Schmidt, and
  Weinstock}}]{Prokopec2004a}
\bibinfo{author}{\bibfnamefont{T.}~\bibnamefont{Prokopec}},
  \bibinfo{author}{\bibfnamefont{M.~G.} \bibnamefont{Schmidt}},
  \bibnamefont{and}
  \bibinfo{author}{\bibfnamefont{S.}~\bibnamefont{Weinstock}},
  \bibinfo{journal}{Ann. Phys. (NY)} \textbf{\bibinfo{volume}{314}},
  \bibinfo{pages}{208} (\bibinfo{year}{2004}{\natexlab{a}}).

\bibitem[{\citenamefont{Prokopec
  et~al.}(2004{\natexlab{b}})\citenamefont{Prokopec, Schmidt, and
  Weinstock}}]{Prokopec2004b}
\bibinfo{author}{\bibfnamefont{T.}~\bibnamefont{Prokopec}},
  \bibinfo{author}{\bibfnamefont{M.~G.} \bibnamefont{Schmidt}},
  \bibnamefont{and}
  \bibinfo{author}{\bibfnamefont{S.}~\bibnamefont{Weinstock}},
  \bibinfo{journal}{Ann. Phys. (NY)} \textbf{\bibinfo{volume}{314}},
  \bibinfo{pages}{267} (\bibinfo{year}{2004}{\natexlab{b}}).

\bibitem[{\citenamefont{Konstandin
  et~al.}(2005{\natexlab{a}})\citenamefont{Konstandin, Prokopec, and
  Schmidt}}]{Konstandin2005a}
\bibinfo{author}{\bibfnamefont{T.}~\bibnamefont{Konstandin}},
  \bibinfo{author}{\bibfnamefont{T.}~\bibnamefont{Prokopec}}, \bibnamefont{and}
  \bibinfo{author}{\bibfnamefont{M.~G.} \bibnamefont{Schmidt}},
  \bibinfo{journal}{Nucl.Phys.} \textbf{\bibinfo{volume}{B716}},
  \bibinfo{pages}{373} (\bibinfo{year}{2005}{\natexlab{a}}).

\bibitem[{\citenamefont{Konstandin
  et~al.}(2005{\natexlab{b}})\citenamefont{Konstandin, Prokopec, Schmidt, and
  Seco}}]{Konstandin2005b}
\bibinfo{author}{\bibfnamefont{T.}~\bibnamefont{Konstandin}},
  \bibinfo{author}{\bibfnamefont{T.}~\bibnamefont{Prokopec}},
  \bibinfo{author}{\bibfnamefont{M.~G.} \bibnamefont{Schmidt}},
  \bibnamefont{and} \bibinfo{author}{\bibfnamefont{M.}~\bibnamefont{Seco}},
  \bibinfo{journal}{eprint hep-ph/0505103}
  (\bibinfo{year}{2005}{\natexlab{b}}).

\bibitem[{\citenamefont{Alford et~al.}(2004)\citenamefont{Alford, Berges, and
  Cheyne}}]{Alford2004a}
\bibinfo{author}{\bibfnamefont{M.}~\bibnamefont{Alford}},
  \bibinfo{author}{\bibfnamefont{J.}~\bibnamefont{Berges}}, \bibnamefont{and}
  \bibinfo{author}{\bibfnamefont{J.~M.} \bibnamefont{Cheyne}},
  \bibinfo{journal}{Phys. Rev. D} \textbf{\bibinfo{volume}{70}},
  \bibinfo{pages}{125002} (\bibinfo{year}{2004}).

\bibitem[{\citenamefont{Aarts and Berges}(2002)}]{Aarts2002a}
\bibinfo{author}{\bibfnamefont{G.}~\bibnamefont{Aarts}} \bibnamefont{and}
  \bibinfo{author}{\bibfnamefont{J.}~\bibnamefont{Berges}},
  \bibinfo{journal}{Phys. Rev. Lett.} \textbf{\bibinfo{volume}{88}},
  \bibinfo{pages}{041603} (\bibinfo{year}{2002}).

\bibitem[{\citenamefont{Petrov et~al.}(2000{\natexlab{a}})\citenamefont{Petrov,
  Holzmann, and Shlyapnikov}}]{Petrov2000a}
\bibinfo{author}{\bibfnamefont{D.~S.} \bibnamefont{Petrov}},
  \bibinfo{author}{\bibfnamefont{M.}~\bibnamefont{Holzmann}}, \bibnamefont{and}
  \bibinfo{author}{\bibfnamefont{G.~V.} \bibnamefont{Shlyapnikov}},
  \bibinfo{journal}{Phys. Rev. Lett.} \textbf{\bibinfo{volume}{84}},
  \bibinfo{pages}{2551} (\bibinfo{year}{2000}{\natexlab{a}}).

\bibitem[{\citenamefont{Petrov et~al.}(2000{\natexlab{b}})\citenamefont{Petrov,
  Shlyapnikov, and Walraven}}]{Petrov2000b}
\bibinfo{author}{\bibfnamefont{D.~S.} \bibnamefont{Petrov}},
  \bibinfo{author}{\bibfnamefont{G.~V.} \bibnamefont{Shlyapnikov}},
  \bibnamefont{and} \bibinfo{author}{\bibfnamefont{J.~T.~M.}
  \bibnamefont{Walraven}}, \bibinfo{journal}{Phys. Rev. Lett.}
  \textbf{\bibinfo{volume}{85}}, \bibinfo{pages}{3745}
  (\bibinfo{year}{2000}{\natexlab{b}}).

\bibitem[{\citenamefont{Andersen et~al.}(2002)\citenamefont{Andersen, {Al\
  Khawaja}, and Stoof}}]{Andersen2002b}
\bibinfo{author}{\bibfnamefont{J.~O.} \bibnamefont{Andersen}},
  \bibinfo{author}{\bibfnamefont{U.}~\bibnamefont{{Al\ Khawaja}}},
  \bibnamefont{and} \bibinfo{author}{\bibfnamefont{H.~T.~C.}
  \bibnamefont{Stoof}}, \bibinfo{journal}{Phys. Rev. Lett.}
  \textbf{\bibinfo{volume}{88}}, \bibinfo{pages}{070407}
  (\bibinfo{year}{2002}).

\bibitem[{\citenamefont{{Al\ Khawaja} et~al.}(2002)\citenamefont{{Al\ Khawaja},
  Andersen, Proukakis, and Stoof}}]{AlKhawaja2002a}
\bibinfo{author}{\bibfnamefont{U.}~\bibnamefont{{Al\ Khawaja}}},
  \bibinfo{author}{\bibfnamefont{J.~O.} \bibnamefont{Andersen}},
  \bibinfo{author}{\bibfnamefont{N.~P.} \bibnamefont{Proukakis}},
  \bibnamefont{and} \bibinfo{author}{\bibfnamefont{H.~T.~C.}
  \bibnamefont{Stoof}}, \bibinfo{journal}{Phys. Rev. A}
  \textbf{\bibinfo{volume}{66}}, \bibinfo{pages}{013615}
  (\bibinfo{year}{2002}).

\bibitem[{\citenamefont{Astrakharchik and Giorgini}(2004)}]{Astrakharchik2004a}
\bibinfo{author}{\bibfnamefont{G.~E.} \bibnamefont{Astrakharchik}}
  \bibnamefont{and} \bibinfo{author}{\bibfnamefont{S.}~\bibnamefont{Giorgini}},
  \bibinfo{journal}{Phys. Rev. A} \textbf{\bibinfo{volume}{68}},
  \bibinfo{pages}{031602(R)} (\bibinfo{year}{2004}).

\bibitem[{\citenamefont{Pilati et~al.}(2005)\citenamefont{Pilati, Boronat,
  Casulleras, and Giorgini}}]{Pilati2005a}
\bibinfo{author}{\bibfnamefont{S.}~\bibnamefont{Pilati}},
  \bibinfo{author}{\bibfnamefont{J.}~\bibnamefont{Boronat}},
  \bibinfo{author}{\bibfnamefont{J.}~\bibnamefont{Casulleras}},
  \bibnamefont{and} \bibinfo{author}{\bibfnamefont{S.}~\bibnamefont{Giorgini}},
  \bibinfo{journal}{Phys. Rev. A} \textbf{\bibinfo{volume}{71}},
  \bibinfo{pages}{023605} (\bibinfo{year}{2005}).

\bibitem[{\citenamefont{K{\"o}hler and Burnett}(2002)}]{Kohler2002a}
\bibinfo{author}{\bibfnamefont{T.}~\bibnamefont{K{\"o}hler}} \bibnamefont{and}
  \bibinfo{author}{\bibfnamefont{K.}~\bibnamefont{Burnett}},
  \bibinfo{journal}{Phys. Rev. A} \textbf{\bibinfo{volume}{65}},
  \bibinfo{pages}{033601} (\bibinfo{year}{2002}).

\bibitem[{\citenamefont{K{\"o}hler et~al.}(2003)\citenamefont{K{\"o}hler,
  Gasenzer, and Burnett}}]{Kohler2003a}
\bibinfo{author}{\bibfnamefont{T.}~\bibnamefont{K{\"o}hler}},
  \bibinfo{author}{\bibfnamefont{T.}~\bibnamefont{Gasenzer}}, \bibnamefont{and}
  \bibinfo{author}{\bibfnamefont{K.}~\bibnamefont{Burnett}},
  \bibinfo{journal}{Phys. Rev. A} \textbf{\bibinfo{volume}{67}},
  \bibinfo{pages}{013601} (\bibinfo{year}{2003}).

\bibitem[{\citenamefont{Berges and Serreau}(2004)}]{Berges2004c}
\bibinfo{author}{\bibfnamefont{J.}~\bibnamefont{Berges}} \bibnamefont{and}
  \bibinfo{author}{\bibfnamefont{J.}~\bibnamefont{Serreau}},
  \bibinfo{journal}{eprint hep-ph/0410330; Proc. Int. Conf. SEWM2004}
  (\bibinfo{year}{2004}).

\bibitem[{\citenamefont{Berges}(2005)}]{Berges2005a}
\bibinfo{author}{\bibfnamefont{J.}~\bibnamefont{Berges}},
  \bibinfo{journal}{eprint hep-ph/0409233; AIP Conf. Proc.}
  \textbf{\bibinfo{volume}{739}}, \bibinfo{pages}{3} (\bibinfo{year}{2005}).

\bibitem[{\citenamefont{Kokkelmans et~al.}(2002)\citenamefont{Kokkelmans,
  Milstein, Chiofalo, Walser, and Holland}}]{Kokkelmans2002b}
\bibinfo{author}{\bibfnamefont{S.~J.~J.~M.~F.} \bibnamefont{Kokkelmans}},
  \bibinfo{author}{\bibfnamefont{J.~N.} \bibnamefont{Milstein}},
  \bibinfo{author}{\bibfnamefont{M.~L.} \bibnamefont{Chiofalo}},
  \bibinfo{author}{\bibfnamefont{R.}~\bibnamefont{Walser}}, \bibnamefont{and}
  \bibinfo{author}{\bibfnamefont{M.~J.} \bibnamefont{Holland}},
  \bibinfo{journal}{Phys. Rev. A} \textbf{\bibinfo{volume}{65}},
  \bibinfo{pages}{053617} (\bibinfo{year}{2002}).

\bibitem[{\citenamefont{Schwinger}(1961)}]{Schwinger1961a}
\bibinfo{author}{\bibfnamefont{J.}~\bibnamefont{Schwinger}},
  \bibinfo{journal}{J. Math. Phys.} \textbf{\bibinfo{volume}{2}},
  \bibinfo{pages}{407} (\bibinfo{year}{1961}).

\bibitem[{\citenamefont{Keldysh}(1964)}]{Keldysh1964a}
\bibinfo{author}{\bibfnamefont{L.~V.} \bibnamefont{Keldysh}},
  \bibinfo{journal}{[Sov. Phys. JETP {\bf 20}, 1018 (1965)] Zh. Eksp. Teor.
  Fiz.} \textbf{\bibinfo{volume}{47}}, \bibinfo{pages}{1515}
  (\bibinfo{year}{1964}).

\bibitem[{\citenamefont{Berges}(2004)}]{Berges2004a}
\bibinfo{author}{\bibfnamefont{J.}~\bibnamefont{Berges}},
  \bibinfo{journal}{Phys. Rev. D} \textbf{\bibinfo{volume}{70}},
  \bibinfo{pages}{105010} (\bibinfo{year}{2004}).

\bibitem[{\citenamefont{Luttinger and Ward}(1960)}]{Luttinger1960a}
\bibinfo{author}{\bibfnamefont{J.~M.} \bibnamefont{Luttinger}}
  \bibnamefont{and} \bibinfo{author}{\bibfnamefont{J.~C.} \bibnamefont{Ward}},
  \bibinfo{journal}{Phys. Rev.} \textbf{\bibinfo{volume}{118}},
  \bibinfo{pages}{1417} (\bibinfo{year}{1960}).

\bibitem[{\citenamefont{Baym}(1962)}]{Baym1962a}
\bibinfo{author}{\bibfnamefont{G.}~\bibnamefont{Baym}}, \bibinfo{journal}{Phys.
  Rev.} \textbf{\bibinfo{volume}{127}}, \bibinfo{pages}{1391}
  (\bibinfo{year}{1962}).

\bibitem[{\citenamefont{Kokkelmans and Holland}(2002)}]{Kokkelmans2002a}
\bibinfo{author}{\bibfnamefont{S.~J.~J.~M.~F.} \bibnamefont{Kokkelmans}}
  \bibnamefont{and} \bibinfo{author}{\bibfnamefont{M.~J.}
  \bibnamefont{Holland}}, \bibinfo{journal}{Phys. Rev. Lett.}
  \textbf{\bibinfo{volume}{89}}, \bibinfo{pages}{180401}
  (\bibinfo{year}{2002}).

\bibitem[{\citenamefont{Yurovsky and Ben-Reuven}(2003)}]{Yurovsky2003a}
\bibinfo{author}{\bibfnamefont{V.~A.} \bibnamefont{Yurovsky}} \bibnamefont{and}
  \bibinfo{author}{\bibfnamefont{A.}~\bibnamefont{Ben-Reuven}},
  \bibinfo{journal}{Phys. Rev. A} \textbf{\bibinfo{volume}{67}},
  \bibinfo{pages}{043611} (\bibinfo{year}{2003}).

\bibitem[{\citenamefont{K{\"o}hler et~al.}(2004)\citenamefont{K{\"o}hler,
  G{\'o}ral, and Gasenzer}}]{Kohler2004a}
\bibinfo{author}{\bibfnamefont{T.}~\bibnamefont{K{\"o}hler}},
  \bibinfo{author}{\bibfnamefont{K.}~\bibnamefont{G{\'o}ral}},
  \bibnamefont{and} \bibinfo{author}{\bibfnamefont{T.}~\bibnamefont{Gasenzer}},
  \bibinfo{journal}{Phys. Rev. A} \textbf{\bibinfo{volume}{70}},
  \bibinfo{pages}{023613} (\bibinfo{year}{2004}).

\bibitem[{\citenamefont{G{\'o}ral et~al.}(2004)\citenamefont{G{\'o}ral,
  K{\"o}hler, Gardiner, Tiesinga, and Julienne}}]{Goral2004a}
\bibinfo{author}{\bibfnamefont{K.}~\bibnamefont{G{\'o}ral}},
  \bibinfo{author}{\bibfnamefont{T.}~\bibnamefont{K{\"o}hler}},
  \bibinfo{author}{\bibfnamefont{S.~A.} \bibnamefont{Gardiner}},
  \bibinfo{author}{\bibfnamefont{E.}~\bibnamefont{Tiesinga}}, \bibnamefont{and}
  \bibinfo{author}{\bibfnamefont{P.~S.} \bibnamefont{Julienne}},
  \bibinfo{journal}{J. Phys. B: At. Mol. Opt. Phys.}
  \textbf{\bibinfo{volume}{37}}, \bibinfo{pages}{3457} (\bibinfo{year}{2004}).

\bibitem[{\citenamefont{Gasenzer}(2004{\natexlab{a}})}]{Gasenzer2004a}
\bibinfo{author}{\bibfnamefont{T.}~\bibnamefont{Gasenzer}},
  \bibinfo{journal}{Phys. Rev. A} \textbf{\bibinfo{volume}{70}},
  \bibinfo{pages}{021603(R)} (\bibinfo{year}{2004}{\natexlab{a}}).

\bibitem[{\citenamefont{Gasenzer}(2004{\natexlab{b}})}]{Gasenzer2004b}
\bibinfo{author}{\bibfnamefont{T.}~\bibnamefont{Gasenzer}},
  \bibinfo{journal}{Phys. Rev. A} \textbf{\bibinfo{volume}{70}},
  \bibinfo{pages}{043618} (\bibinfo{year}{2004}{\natexlab{b}}).

\bibitem[{\citenamefont{Bettencourt and Wetterich}(1998)}]{Bettencourt1998a}
\bibinfo{author}{\bibfnamefont{L.~M.} \bibnamefont{Bettencourt}}
  \bibnamefont{and}
  \bibinfo{author}{\bibfnamefont{C.}~\bibnamefont{Wetterich}},
  \bibinfo{journal}{Phys.Lett.} \textbf{\bibinfo{volume}{B430}},
  \bibinfo{pages}{140} (\bibinfo{year}{1998}).

\bibitem[{\citenamefont{Aarts et~al.}(2000)\citenamefont{Aarts, Bonini, and
  Wetterich}}]{Aarts2000a}
\bibinfo{author}{\bibfnamefont{G.}~\bibnamefont{Aarts}},
  \bibinfo{author}{\bibfnamefont{G.~F.} \bibnamefont{Bonini}},
  \bibnamefont{and}
  \bibinfo{author}{\bibfnamefont{C.}~\bibnamefont{Wetterich}},
  \bibinfo{journal}{Phys. Rev. D} \textbf{\bibinfo{volume}{63}},
  \bibinfo{pages}{025012} (\bibinfo{year}{2000}).

\bibitem[{\citenamefont{Mihaila et~al.}(1997)\citenamefont{Mihaila, Dawson, and
  Cooper}}]{Mihaila1997a}
\bibinfo{author}{\bibfnamefont{B.}~\bibnamefont{Mihaila}},
  \bibinfo{author}{\bibfnamefont{J.~F.} \bibnamefont{Dawson}},
  \bibnamefont{and} \bibinfo{author}{\bibfnamefont{F.}~\bibnamefont{Cooper}},
  \bibinfo{journal}{Phys. Rev. D} \textbf{\bibinfo{volume}{56}},
  \bibinfo{pages}{5400} (\bibinfo{year}{1997}).

\bibitem[{\citenamefont{Ryzhov and Yaffe}(2000)}]{Ryzhov2000a}
\bibinfo{author}{\bibfnamefont{A.~V.} \bibnamefont{Ryzhov}} \bibnamefont{and}
  \bibinfo{author}{\bibfnamefont{L.~G.} \bibnamefont{Yaffe}},
  \bibinfo{journal}{Phys. Rev. D} \textbf{\bibinfo{volume}{62}},
  \bibinfo{pages}{125003} (\bibinfo{year}{2000}).

\end{thebibliography}

\end{document}